%% file: paper.tex
\documentclass[conference, compsoc]{IEEEtran}
\IEEEoverridecommandlockouts
\usepackage{subfigure}
\usepackage{caption}
\usepackage{hyperref}

\usepackage{amsmath}
\usepackage{booktabs}
\usepackage{graphicx}
\usepackage{multirow}
\usepackage{courier}
\usepackage{multirow}
\usepackage{acronym}
\usepackage{color}

\begin{document}

\title{Distributed Hybrid Simulation of the\\Internet of Things and Smart Territories\footnotemark}

\author{\IEEEauthorblockN{Gabriele D'Angelo, Stefano Ferretti, Vittorio Ghini}
\IEEEauthorblockA{Department of Computer Science and Engineering (DISI), University of Bologna, Italy\\
\{g.dangelo, s.ferretti, vittorio.ghini\}@unibo.it}
}

\maketitle

\footnotetext{The publisher version of this paper is available at \url{https://doi.org/10.1002/cpe.4370}.
\textbf{{\color{red} This is the pre-peer reviewed version of the following
article: ``Gabriele D'Angelo, Stefano Ferretti, Vittorio Ghini. Distributed Hybrid Simulation of the Internet of Things and Smart Territories. Concurrency and Computation: Practice and Experience (Wiley), vol. 30, issue 9 (May 2018)''.}}}

\begin{abstract}
This paper deals with the use of hybrid simulation to build and compose heterogeneous simulation scenarios that can be proficiently exploited to model and represent the Internet of Things (IoT). Hybrid simulation is a methodology that combines multiple modalities of modeling/simulation. Complex scenarios are decomposed into simpler ones, each one being simulated through a specific simulation strategy. All these simulation building blocks are then synchronized and coordinated. This simulation methodology is an ideal one to represent IoT setups, which are usually very demanding, due to the heterogeneity of possible scenarios arising from the massive deployment of an enormous amount of sensors and devices. We present a use case concerned with the distributed simulation of smart territories, a novel view of decentralized geographical spaces that, thanks to the use of IoT, builds ICT services to manage resources in a way that is sustainable and not harmful to the environment. Three different simulation models are combined together, namely, an adaptive agent-based parallel and distributed simulator, an OMNeT++ based discrete event simulator and a script-language simulator based on MATLAB. Results from a performance analysis confirm the viability of using hybrid simulation to model complex IoT scenarios.\\
\end{abstract}


\input{sec_introduction.tex}

\input{sec_background_revised.tex}

\input{sec_multilevelsimulation.tex}

\input{sec_casestudy.tex}

\input{sec_perf.tex}

\input{sec_conclusions.tex}

\small{
\bibliographystyle{abbrv}
\bibliography{paper}  
}

\end{document}

%% file: sec_introduction.tex
\section{Introduction}
\label{sec:intro}
The Internet of Things (IoT) is a recent term coined to identify the rapidly growing multitude of sensors and mobile users' terminals connected to the Internet. 
The term ``Internet of Things'' is employed in several different domains, often exploited as a buzzword in many scientific and technological domains.
The term was originally coined by K.~Ashton to refer to a system of ubiquitous sensors connecting the physical world to the Internet.
In \cite{vermesan2011internet}, IoT is defined as a ``\textit{dynamic global network infrastructure with self configuring capabilities based on standard and interoperable communication protocols where physical and virtual ‘things’ have identities, physical attributes, and virtual personalities and use intelligent interfaces, and are seamlessly integrated into the information network}''. In documents from the European Commission, the IoT is seen as a general ``\textit{evolution of the Internet from a network of interconnected computers to a network of interconnected objects}''.

The interactions of these things and the data they produce (or sense) might be somehow utilized and managed to offer novel services in smart cities and territories in general~\cite{iot_survey,Atzori:2010,gda-hpcs-16,gda-simpat-2017}.
Examples of services can be found in a broad range of applications areas. 
As an example, smart transportation systems can be built where cars, equipped with propers sensors, and static sensors deployed in the streets interact to form a crowd-sensing platform, whose data can be exploited, for instance, to reduce congestion, optimize emergency services response times, lower fuel usage, reduce pollution, offer parking availability information, promote the development of smart safety systems, etc. 
But in general, when the big data coming from these sensors are released as open data, this information can be used in a variety of applications, which might be quite different from the original applications they have been generated for.
This leads to the possible development of novel (social) applications based on crowd- sourced and sensed data~\cite{PrandiFMS15}.
Not only, the interactions among things might lead to novel opportunistic and self configuring services in rural territories, exploiting both infrastructured and infrastructure-less communication networks~\cite{Atzori:2010,Ferretti2013481,smartshires,smartshires_abps}.

In many cases, these devices are equipped with a very little amount of memory and computational power, constrained software and administration capabilities, e.g.,~limited administration utilities and few system updates.
Being able to understand and to simulate the IoT will soon become essential. The complex networks obtained by the interaction of IoT devices are hard to design and to manage. In real deployment scenarios, many configurations of IoT networks are possible. Devices’ connectivity is influenced by their geographical location, communication and network capabilities, device distribution.

Thus, modeling an IoT environment can result in a difficult task, due to the heterogeneous possible scenarios. IoT simulation is necessary for both quantitative and qualitative aspects. To name a few issues: capacity planning, what-if simulation and analysis, proactive management and support for many specific security-related evaluations. For instance, through the modeling and simulation of the IoT, it is possible to understand the amount of sensors to deploy in a given area and identify their optimal location. It is possible to simulate specific, critical scenarios, which cannot be tested in real situations and understand the adequacy of the specific configuration of things in a given area. 

The scale of the IoT is the main problem in the usage of existing simulation tools. Traditional approaches (that are single CPU-based) are often unable to scale to the number of nodes (and level of detail) required by the IoT. 

This paper deals with the use of hybrid simulation to build and compose heterogeneous simulation scenarios that can be proficiently exploited to model IoT environments. Hybrid simulation is a field of modeling and simulation, which comprises several methodologies proposed by different authors in different domains~\cite{Eldabi:2016:HSH:3042094.3042274}. In our view, hybrid simulation combines multiple modalities of different modeling and simulation techniques in order to build sophisticated tools for the analysis of any kind of systems. 
Complex scenarios can be decomposed into simpler ones, each one being simulated through a specific simulation strategy and possibly a domain-specific simulator. Hence, different levels of detail and types of simulation can be exploited to model peculiar aspects of the system being simulated. Synchronization, coordination and interaction of these simulation building blocks are needed to realize sophisticated hybrid simulation tools, avoid causality errors and guarantee the expected runtime performance of the combined simulator. This novel simulation technique allows for high scalability, especially when combined with adaptive Parallel and Distributed Simulation (PADS) approaches. 

We claim that hybrid simulation can be proficiently employed to effectively simulate the IoT. As mentioned, the range of issues that need to be accounted for in IoT is wide. Classic simulation approaches could not be able to consider all these aspects within a single simulator. Rather, hybrid approaches allow creating complex simulation models that consider all relevant issues, when needed. Multi-level techniques can be employed so that a simulator can trigger, only when needed, the execution of a finer-grained simulator, able to mimic with high detail certain issues of interest.

We demonstrate the validity of the proposed approach by focusing on the simulation of ``smart territories''~\cite{smartshires,smartshires_abps}.
This is a novel view of devising smart services over urban and decentralized environments. Indeed, in these last months/years focus has been given on the development of smart cities services, i.e.~a set of strategies aiming at improving and optimizing services offered to citizens living in metropolitan areas.
As a matter of fact, the possibility to offer services for territorial districts with low population density is an almost ignored problem. 
There is a need to devise smart, cheap and sustainable services in decentralized geographical spaces, without the need of costly (communication) infrastructures. Such services would make good use of a deployment of cheap sensors in these areas, together with ad-hoc configurations of mobile devices.
We show that the design and configuration of smart services in (decentralized) territories impose the simulation of wide area networks; however, in certain cases a highly detailed simulation is required. This need for scalability and high level of detail can be reached by resorting to properly configured hybrid simulation techniques.
An advantage of this approach is that the detailed (and thus, more costly) simulation can be performed only when needed, in a limited simulated area, only for the needed time interval of the simulation.

As a use case, we focus on the simulation of a ``smart market scenario''. We assume that users can subscribe their interest for certain kinds of products to a service that informs them upon availability of interesting events, sales, availability in markets within their neighborhood. Thus, users can decide to order products or move to the market location for some shopping. On their way, they need to be guided to find the specific place and then the products.

The developed hybrid simulator is composed of three distributed, interacting simulators, executing at two different levels of detail. 
A coarse level simulates the whole smart territory, where different actors produce products, subscribe their interests, move towards different geographical areas. This has been implemented using a discrete event simulator following an agent-based modeling approach, equipped with PADS capabilities called GAIA/ART\`IS~\cite{gda-simpat-2017-part}.
The arrival of customers (through a transportation system) to the location, and the parking deployment strategies in the neighborhood of the market have been implemented as a simulator that exploits the equation-based ADVISOR tool~\cite{advisor} that is implemented on top of MATLAB.
Finally, the interactions of pedestrian users within the smart market have been developed through an instance of an OMNeT++ discrete event simulation; this simulator considers wireless communication issues, fine-grained interactions and movements. 
We provide an experimental evaluation of the proposed distributed hybrid simulator and of its simulation components, that confirms the viability of the proposed approach.

The remainder of this paper is organized as follows.
Section \ref{sec:background} describes the background and state of the art about simulation and IoT/Smart-Territories.  
The proposed approach, based on hybrid simulation, is introduced in Section \ref{sec:hybridsimulation}. In Section \ref{sec:casestudy}, this approach is applied to a ``smart shires'' case study. Section \ref{sec:perf} presents some results on a performance assessment on the hybrid simulator we utilized to model the use case. Finally, Section \ref{sec:conc} provides some concluding remarks.

%% file: sec_background_revised.tex
\section{Background and State of the Art}
\label{sec:background}

In this section we provide the background, needed in the rest of the work, and an overview of the state of the art. In particular, we first provide a discussion on Discrete Event Simulation (DES), followed by the background of modern techniques, which are based on the parallelization and distribution of DES, i.e.,~PADS. Then, we briefly outline the possibility of employing adaptive schemes to migrate distributed simulation entities (e.g.,~agents) and cluster them together, so that interacting simulation entities are executed into the same logical processes. We then provide a general introduction on hybrid modeling and simulation, which will be discussed in more detail in the next section.

Finally, we introduce the background related to the use case which is considered in detail in this paper, i.e.,~Internet of Things and smart territories.

\subsection{Discrete Event Simulation}
In a computer simulation, a process models the behavior of some other system over time~\cite{FUJ00}. In some cases, the simulated system is real but more often it has yet to be designed or implemented. In practice, simulation is about methodologies and techniques that are needed for the performance evaluation of complex systems.

The motivations behind the use of simulation are many. To name a few: cost reasons, testing on the real system is too dangerous, many different solutions must to be evaluated to support the system design (i.e.~dimensioning and tuning). Due to the increasing complexity in the systems to be built, simulation is used more and more often. 

Discrete Event Simulation (DES)~\cite{Law:1999:SMA:554952} is one of the many simulation paradigms that have been proposed. With respect to other approaches, it has good expressiveness and it is quite easy to use. A DES is represented by a simulated model (that is implemented using a set of state variables) and its evolution (that is represented by a sequence of events processed in chronological order). Each event occurs at a given instant in time and represents a change in the simulated model state. This means that the whole evolution of the simulated system is obtained through the execution of an ordered sequence of events that are: created, stored and processed. 
For example, the events in the simulation of Vehicular Ad Hoc Networks are the updates of the cars positions and the transmission of data packets. At the basics, a DES is a set of state variables (i.e.~describing the modeled system), an event list (i.e.~the pending events that will be processed for evolving the simulated state) and global clock (i.e.~the simulation time)~\cite{Law:1999:SMA:554952}. Each event is tagged by a timestamp that specifies the simulated time at which it has to be processed.
DES models usually embody some kind of randomness in the generation of events. This means that, in order to have a comprehensive and clear understanding of the simulated system, during the experimental assessment the simulation runs must be repeated several times. This allows performing some statistical analysis on the obtained metrics.

In a sequential (i.e.~monolithic) simulation, a single Physical Execution Unit (PEU), for example a CPU core, is in charge of creating new events, updating the pending event list and processing the events in timestamp order. 
In other words, a program executed on a single CPU core manages the whole simulated model and its evolution. 
This approach is simple and easy to implement but it has some drawbacks. Among others, the simulation scalability both in terms of execution time (to complete the simulations runs) and size of the system that can be represented~\cite{1668384}.

\subsection{Parallel DES and PADS}
\label{PDES_PADS}
As an alternative, the tasks described above can be parallelized using a set of interconnected PEUs (e.g.~CPU cores, CPUs or hosts). This approach is called Parallel Discrete Event Simulation (PDES)~\cite{Fujimoto:1989:PDE:76738.76741}. In this case, very large and complex models can be represented and executed since each PEU is only in charge of a part of the simulation model. That is, each PEU manages a local pending events list and some events are delivered by means of messages to remote PEUs. In addition, the PEUs must run a synchronization algorithm to guarantee the correct simulation execution. In many cases, a PDES approach can speedup the simulation execution, at the cost of a more complex implementation and setup of the simulator.

A Parallel and Distributed Simulation (PADS) is a simulation that is run on more than one processor~\cite{perumalla2007}. There are many good reasons to rely on this approach, among them: execution speed, model scalability, interoperability and composability purposes (e.g.~to integrate different off-the-shelf simulators and to compose many already existing simulation models in a new simulator)~\cite{FUJ00}.

\begin{figure}[ht]
\centering
\includegraphics[width=7.0cm]{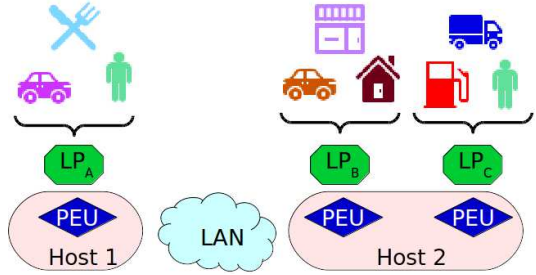}
\caption{Parallel and Distributed Simulation: model partitioning.}
\label{fig:pads-model}
\end{figure}

With respect to a monolithic simulation, a PADS lacks a global model state. That is, a single representation of the simulated model is missing. In fact, each PEU in the PADS manages only a part of the simulated model. Following the PADS terminology, the model components executed on top of each PEU are called Logical Processes (LPs)~\cite{gda-hpcs-11}. As shown in Figure~\ref{fig:pads-model}, a PADS is obtained through the interaction among LPs; each LP deals with the evolution of a part of the simulated model and interacts with other LPs (for synchronization and data distribution)~\cite{FUJ00}.

The performance of the network that interconnects the LPs has a strong effect on the PADS design and the simulator execution speed. When the LPs are run on PEUs interconnected by a shared memory, then it is called parallel simulation. 
Conversely, loosely coupled LPs, i.e.~where every LP is an autonomous independent system connected to other LPs via some network infrastructure, are referred as distributed simulation~\cite{Boukerche:2000}.
More often, the execution architecture used to run PADS are a mix of parallel and distributed PEUs~\cite{gda-simpat-2014}.

In short, the main issues in a PADS are:
\begin{itemize}
	\item the simulated model is partitioned in a set of LPs~\cite{bagrodia98}. The \textbf{partitioning} is a complex task since it must be done by considering both the minimization of the network communication (among LPs) and the load balancing in the parallel/distributed execution architecture;
	\item the results obtained by the PADS are correct only if they are exactly the same given by the sequential simulator.	This can happen only if there is a \textbf{synchronization} algorithm that properly coordinates the LPs evolution;
	\item each LP generates updates (events) that are possibly relevant for parts of the simulated model in other LPs. For performance reasons, broadcasting all events is not feasible. \textbf{Data distribution} is about the efficient delivery of state updates and it is often based on a publish-subscribe approach~\cite{Jun:2002:ESM:564062.564074}.
\end{itemize}

Implementing a PDES using PADS means that each event generated by a LP must be sent to other interested LPs. As mentioned, LPs associate timestamps to generated events. Then, events are encapsulated into messages to be transmitted for the inter-LP delivery. 

As defined by Lamport: ``two events are in causal order if one of them can have some consequences on the other''~\cite{Lamport1978}. Clearly, to get a correct simulation execution, the causal order of events must not being violated. This is easy in a monolithic simulation but it is complex in parallel and distributed architectures due to the different execution speeds of each PEU and the network delays. In a PADS, to guarantee that all events are executed in non-decreasing timestamp order, the LPs have to run a synchronization algorithm. The synchronization can be handled in many ways but the main approaches are the following:
\begin{itemize}
	\item \emph{time-stepped}: the simulated time is divided in timesteps	of fixed-size. The simulation model is updated at every timestep and the lower bound to the flight time for interactions between the model components is the size of the timestep. When a LP completes the tasks for the current timestep, it broadcasts to all the other LPs an End-Of-Step (EOS) message and then waits the EOS messages from all other LPs before proceeding to the next timestep~\cite{1261535};
	\item \emph{conservative}: in this approach the causality errors are prevented. That is, before processing each event, it is checked if the event is ``safe'' or not (with respect to the causality constraint). If the event is tagged as safe by the synchronization algorithm then it can be processed. Otherwise, the LP must stop processing while waiting for more events (or better information about the safety of events). This safety check can be implemented in many different ways, a widely used algorithm is the Chandy-Misra-Briant~\cite{misra86};
	\item \emph{optimistic}: in this case the events are processed by the LPs in receiving order. This means that, very likely, the causality order will be violated. In fact, when a violation is found by the synchronization algorithm, the LP that has found it implements a roll-back to the (most recent) previous state that is correct. Furthermore, it propagates the roll-back to all the other LPs that have been affected by the violation~\cite{timewarp}. In this way, the whole PADS goes back to the most recent globally correct simulation state and it starts again processing the events.
\end{itemize}

\subsection{Adaptive PADS}
As described before, the partitioning of the simulated model in PADS is a complex task. Over the years, many static and dynamic approaches have been proposed to automate and enhance the partitioning of parallel and distributed simulations. The most relevant partitioning approaches have been discussed in \cite{gda-simpat-2017-part}. In the same paper, we have proposed an approach in which the simulated model is represented by a multi-agent system. The simulated model is partitioned in small model components (also called Simulated Entities, SEs) and the model evolution is obtained through the exchange of interactions among SEs. In this way, the LPs are containers of SEs and it is possible to move (migrate) a SE from one LP to another. This permits to avoid the static partitioning of the simulated model and to adaptively reallocate the SEs for better computational and communication load balancing. In many cases, this leads to a speed up in the simulation execution and enhanced scalability. This adaptive PADS approach is implemented in the GAIA/ART\`IS simulator~\cite{pads}.

\subsection{Agent-based Simulation}
A particular type of simulation is the Agent Based Modeling and Simulation (ABMS)~\cite{north}. In ABMS, the simulated entities are called agents. Agents can represent any actor entity within a simulation. The model specifies the behavior of each agent (i.e.,~microscale model). Such a behavior is usually influenced by the information that agents obtain from the environment. Thus, events generated in the environment have an impact on the agents states, and the whole simulation evolves based on these interactions among agents and the environment. Typically, the simulation model specifies one or more classes of agents; each agent of a certain class executes the same behavior procedure. The variety of the possible outcomes of a simulation is thus due to some kind of randomness (introduced in the specification of such a behavior) and the different interactions and situations each agent is involved in.

Over the years, a huge number of agent-based models have been developed and it has been demonstrated that in some circumstances ABMS offers some advantages with respect to other approaches~\cite{Macal:2009:AMS:1995456.1995474}.

In massively populated simulation environments (such as in the IoT, where each thing is an agent) certain problems may become too big to be solved in a serial computing environment. The memory and computational requirements cannot be supplied by a single CPU. Hence, parallel and distributed agent-based simulation comes into the picture.
However, as in the previous scenarios, parallelization and distribution do not come for free. It is important to distribute the model as evenly as possible, try to minimize message passing requirements between LPs, synchronize all the LPs to ensure a correct and consistent simulation. 

There are examples of classic ABMS tools that provide some kind of extensions to distribute the computation. For instance, D-MASON is the distributed version of the popular MASON tool, a Java-based ABMS~\cite{Scarano2014}. The ABM++ framework is a tool to implement agent-based models to be deployed on distributed memory Linux clusters \cite{abm++}. The framework provides the necessary functionality to allow applications to run on distributed architectures. The usage of distributed discrete-event simulation techniques for the simulation of multi-agent systems~\cite{910853} involves some common problems of PADS such as the lack of a shared state, the interest management and the load balancing of the distributed execution architecture. Other aspects, such as the ``spheres of influence'' are from agent-based modeling but, in this case, must be addressed in a distributed setup. Some efforts have been done to implement the distributed simulation of agent-based systems using the High Level Architecture (IEEE 1516) standard~\cite{Lees:2007:DSA:1243991.1243992}. The simulation of systems with million of agents has led to the development of frameworks that are able to run on GPUs~\cite{flamegpu} and clusters composed of multiple GPUs and multi-core processors~\cite{Aaby:2010:ESA:1808143.1808181}.

Other examples in literature are works on specific implementations of distributed agent-based simulation tools to be used in specific use cases. Just to mention a few, LUNES is a parallel and distributed agent-based Large Unstructured NEtwork Simulator, which allows to simulate complex networks composed of a high number of nodes~\cite{gda-mospas-11}. In \cite{Parker2011}, a distributed platform, termed Global-Scale Agent Model (GSAM), is presented. It is thought to build agent-based epidemic models for the simulation of disease outbreaks. In \cite{Collier2015}, a case study is discussed on the use of the OpenMP toolkit and MPI to distribute a large-scale epidemiologic agent-based model.

\subsection{Hybrid Modeling and Simulation}
Common techniques employed in simulation, being DES, PADS, Monte Carlo, agent-based, equation based modeling techniques and so on, are typically applied in isolation one from another. Hybrid simulation rests on the idea of employing different simulation techniques together. Even if there is no a cohesive and overall accepted definition for hybrid simulation~\cite{Eldabi:2016:HSH:3042094.3042274}, it is clear that the idea of mixing analytic and simulation models is not new~\cite{717204, 10.2307/170837} and, in the past, it has been already implemented in a few simulation packages~\cite{Mosterman1999}. In our view, a definition of hybrid simulation should consider hybrid models that are based on two or more simulation models such as linking discrete event simulation (DES) with either system dynamics (SD) or agent based (ABS)~\cite{Eldabi:2016:HSH:3042094.3042274} but also other analytical models (e.g.,~continuous simulation). It is worth noting that our approach is partially different from \cite{717204}, in which hybrid models and hybrid modeling have two very different definitions. Indeed, in \cite{7020017} a conceptual analysis of hybrid simulation typologies is provided. In that paper, a distinction is made between ``hybrid simulation'' and ``hybrid modeling and simulation (M\&S)''. The former is referred as the application of multiple techniques in the model implementation stage of a simulation. And it is true that the majority of papers referring to hybrid simulation focus on the model implementation phase of a simulation study \cite{6721545,Eldabi:2016:HSH:3042094.3042274,Mosterman1999,10.2307/170837}. The latter deploys the use of inter-disciplinary methods and comprises several stages other than implementation, concerned with problem formulation stage/conceptual modeling, data collection, validation and verification, model execution, data analysis etc.~\cite{6465011}.

Even if, in the last years, hybrid simulation has been mostly used in specific fields such as healthcare~\cite{6721545}, other fields could benefit from its application. For example, production planning~\cite{Byrne1999305}, operations research~\cite{7020017}, biological systems~\cite{KiehlMS04} and computer systems~\cite{Schwetman:1978:HSM:359588.359594}.

\subsection{Internet of Things and Smart-Territories}
In the previous section, we already mentioned the trend towards the design of novel services, built by interconnecting various heterogeneous devices, deployed in geographical areas~\cite{Petrolo:2014}. 
Smart services can be built that exploit data collected by things deployed in a geographical location such as a metropolitan, crowded area or even a rural country-side~\cite{smartshires,smartshires_abps}. 
In fact, sensors are relatively cheap in terms of cost. Thus, their massive deployment is feasible both in populated centres and in more decentralized areas~\cite{smartshires}. Such sensors can be interconnected to form a sensor network. 
Data sensed by the sensors' devices can be disseminated and collected by some information processing system, treated as open data and managed through a context-aware data distribution service, to be used by applications~\cite{iot_survey}.

Clearly enough, the larger the audience for a service the higher the amount of resources that can be possibly devoted to support the service. Conversely, the more decentralized the location the higher the need for the devised solutions to be cheap and sustainable.
In particular, the presence of an available infrastructured wireless communication network, typically present in urban areas, promotes the use of software solutions where the (big) data produced or sensed by things are collected, aggregated and exploited by intelligent services hosted in cloud (or fog) computing architectures. 
These smart services can integrate such data with crowd-sensed and crowd-sourced data coming from mobile users' terminals~\cite{Khan:2012,Lea:2014,PereraZCG14}. 

Conversely, in a decentralized (rural) area, self-configuring opportunistic solutions might be preferred, possibly not strictly dependent on the presence of a classic networking infrastructure~\cite{gda-simpat-2017}. 
Whatever the distributed software solution, the main idea is that the IoT represents the substrate to build smart services to foster the creation of ``smart territories''.

The complexity of the possible scenarios coming from this picture suggests that effective simulation tools are needed. These simulation tools must take into consideration issues concerned with complex networks, aspects typical of pervasive computing and low-level details concerned with wireless communications. In the next sections, we will discuss existing methods and viable strategies to ensure scalability of the simulation, without introducing oversimplifications and inaccuracies, due to the lack of the level of detail.

\subsubsection{Simulation of the Internet of Things}
The design of complex IoT setups requires the support of large scale testbeds or the usage of scalable simulation tools. 
In the case of simulation, the number of nodes in the scenario and the level of detail required by the interaction among nodes are key elements for the scalability of the simulator. 
Put in other words, a simulator of the IoT should be evaluated based on its ability to scale and to offer methods to simulate diverse, highly detailed aspects related to the interaction among things.
This is evident from the need of simulating a massive amount of things, and the diverse application scenarios and issues to be simulated, ranging from smart services in smart cities, (social) applications based on crowd- sourced and sensed data, intelligent transportation systems, up to opportunistic and self configuring communications in rural territories~\cite{iot_survey,smartshires,Ferretti2013481}.

In~\cite{6069710}, the authors identify the requirements for the next generation of IoT experimental facilities, they discuss some drawbacks of simulation-based approaches and provide a survey of existing testbeds (some of them also supporting co-simulation). An approach based on the federation of testbeds is possible but it has many drawbacks. 
In most cases, existing network simulators are inadequate for the scale and level of detail required by IoT models.

SimIoT is a new simulator described in~\cite{6844677} in which the back-end operations are executed in a cloud environment for better performance. The use case proposed in the paper is a health monitoring system for emergency situations in which short range and wireless communication devices are used to monitor the health of patients. The preliminary performance evaluation is based on 160 identical jobs submitted by 16 IoT devices.

In~\cite{6664581} the massive-scale of many IoT deployments is considered. In this case, the authors firstly present a survey of large-scale simulators and emulators and then they propose MAMMotH, a software architecture based on emulation. To the best of our knowledge the development of MAMMotH stopped in 2013.

Brambilla et al. propose to integrate the DEUS general-purpose discrete event simulation with the domain specific simulators Cooja and ns-3 for the study of large-scale IoT scenarios in urban environments~\cite{Brambilla:2014:SPL:2694768.2694780}. In this case, the performance evaluation is based on 6 scenarios with up to $200000$ sensors, $400$ hubs and $25000$ vehicles. The execution time with respect to the number of events shows a quite good scalability. On the other hand, to the best of our knowledge, the DEUS simulator has a monolithic architecture and it is implemented in Java.

In~\cite{6418824}, the authors propose an IoT-based smart home system, in which the performance evaluation is based on different simulation methods such as Monte Carlo.

DPWSim is a simulation toolkit that supports the modeling of the OASIS standard ``Devices Profile for Web Services'' (DPWS)~\cite{6803226}. Its main goal is to provide a cross-platform and easy-to-use assessment of DPWS devices and protocols. In other words, it is not designed for very large-scale setups.

The approach followed in \cite{Brumbulli2016} uses a model-driven simulation (based on the standard language SDL) to describe the IoT scenario.  Starting from this, an automatic code generation transforms the description into an executable simulation model for the ns-3 network simulator.

An interesting approach is proposed in \cite{paper+kirsche-13:iot-simulation}. The author proposes a hybrid simulation environment in which the Cooja-based simulations (i.e.~system level) are integrated with a domain specific network simulator (i.e.~OMNeT++).

Finally, in~\cite{gda-simpat-2017}, the authors present a preliminary approach to the design and performance evaluation of large scale IoT deployments. More specifically, a multi-level simulator composed of two DES-based simulators is proposed and evaluated. The proposed architecture is extended (i.e.~hybrid) and revised in this paper to address more general IoT scenarios.

\subsubsection{Hybrid Simulation of the Internet of Things and Smart-Territories}

As concerns the use of IoT to build efficient services for making ``smarter'' territories, from a simulation point of view
there are many requirements that the simulation tool must provide. Above all, the main issue is scalability, both in terms of amount of modeled entities and granularity of events. Even a small size smart territory will be composed of thousands of interconnected devices. Many of them will be mobile and each with very specific behavior and technical characteristics~\cite{smartshires}. 
If a proactive approach is needed (e.g.~simulation in the loop), in order to perform ``what-if analysis'' during the management of the deployed architecture, then the simulator should be able to run in (almost) real-time, at least with average size model instances. 

We claim that a multi-level simulation is needed in order to simulate a smart territory scenario with a reasonable IoT model. In fact, running the whole model at the highest level of detail is unfeasible. 
Indeed, imagine the computational efforts required to simulate all technical aspects related to each single sensor and device deployed in a territory of interest, together with all the details related to communication issues and the network infrastructure. Not only the amount of these things to simulate make it unfeasible, but it would be quite complex to model all these characteristics with a single simulation tool. In essence, the main problems relate to scalability and expressiveness of the simulation models.
A better approach is to bind different simulators together, each one running at its appropriate level of detail and with specific characteristics of the domain to be simulated (e.g.~mobility models, wireless/wired communications and so on). A further benefit is that each simulator can be activated on demand by the simulator orchestrator (being either a human or a computer program), only when needed.
We will discuss this approach in the next section.

Agent-based simulation is a perfect tool to create models that mimic urban systems in general~\cite{Karnouskos}. 
According to the agent-based simulation methodology, one can specify a general (pseudo-random) behavior of a simulated entity (agent), once. Then, multiple instances (agents) of these entities can be deployed into the simulated world, each one acting in a similar (but different) way. This approach allows studying the emergent behavior of a global system, by specifying single and local interactions. In the context of urban systems, it suffices to define the behavior of one or more types of entities, e.g.,~mobile users, vehicles, and then create and deploy several instances to study the properties of the global urban system. In other terms, agent-based simulation is a powerful tool that offers bottom-up understandings to complex consequences in decision-making and problem-solving processes, as opposed to traditional aggregated modeling approaches~\cite{Chen2012166}.

Agent-based simulation, together with land-use transport interaction model and cellular automata are applicable in planning support systems. These models can be applied at different time scales, such as short-term modeling, e.g. diurnal patterns in cities, and long-term models for exploring change through strategic planning.
Tools such as MASON~\cite{Luke:2005} and SUMO~\cite{SUMO2012} allow simulating moving entities (e.g.~mobile users of vehicles) that can interact with static ones. These tools have been successfully exploited to study intelligent traffic control systems~\cite{bauza,kerekes,Wegener:2008,e16052384}, mobile applications that resort to crowdsensed data~\cite{PrandiFMS15} and so on. The main problem of these approaches is that, due to their nature, they do not allow creating massive scenarios, with many interconnections.

CupCarbon is a multi-agent and discrete event, smart-city and Internet of Things Wireless Sensor Network (SCI-WSN) simulator~\cite{Mehdi:2014}. Its allows designing, visualizing and validating distributed algorithms in a network. It employs the OpenStreetMap framework to deploy sensors directly on the map. The main goal of this tool is to help trainers to explain the basic concepts and how sensor networks work and it can help scientists to test their wireless topologies, protocols, etc. The main problem of scalability remains.

Moreover, it is worth mentioning that there are a number of image and 3D based simulators, such as CanVis, Second Life, Suicidator City Generator, Blended Cities. Among them, 
UrbanSim is a software-based simulation for urban areas, with tools for examining the interplay between land
use, transportation, and policy~\cite{urbansim}. It is intended for use by Metropolitan Planning Organizations and others needing
to interface existing travel models with new land use forecasting and analysis capabilities.
UrbanSim does not focus on scenario development, as most of these tools do, but rather
on understanding the consequences of certain scenarios on urban communities.
However, typically these tools do not cope with issues concerned with wireless communications and pervasive computing, which are the keywords related to the IoT world.

%% file: sec_multilevelsimulation.tex
\section{Hybrid Simulation of the Internet of Things}
\label{sec:hybridsimulation}

Since many IoT models are composed of a very large number of nodes, the usage of a single simulator that embodies all the possible aspects of the simulated world would be quite complex~\cite{gda-hpcs-17,gda-simpat-2017,6069710,6664581,paper+kirsche-13:iot-simulation}. For instance, the use of a fine grained simulation model, according to which all low level details of devices and their interconnections are considered, might lead to scalability problems. 
In essence, a monolithic simulator, that handles all the nodes in the IoT and implements a fine grained level of detail, is unable to provide the simulation results in an acceptable amount of time~\cite{gda-mospas-11,misra86,perumalla2007}. This forces the simulationist to employ a limited amount of simulated entities. 

It is also worth noticing that the use of a monolithic simulator requires that such a simulator has already been built, or it should be implemented from scratch. This might somehow hinder code re-utilization. A modular approach may have some benefits.

On the other hand, using a fine-grained model, for the whole simulation, in a PADS over High Performance Computing execution platforms, might be quite costly. In fact, a more detailed simulation corresponds to higher computations and a higher amount of messages to be exchanged among logical processes. This might also correspond to higher costs related to resource provisioning and maintenance. 
Moreover, it is important to notice that in certain situations the price paid for such fine-grained simulation might be useless, i.e.,~sometimes all the details might be useless (for example, considering wireless networking issues when no communications have to be simulated);  thus, it might be preferable to use a higher level of abstraction.
However, in general, reducing the level of detail in the simulation model might lead to misleading (or wrong) simulation results due to the excessive amount of details removed from the simulated model.

Thus, in this work, we focus on the use of a hybrid simulation approach, coupled with PADS, for large scale IoT setups. That is, a simulation in which multiple simulation models are linked together~\cite{magne2000towards}, each one with a specific task, possibly working at a different level of detail.

Under the implementation viewpoint, this means using a ``high level'' (preferably PADS) simulator, that works at a coarse grained level of detail and that coordinates the execution of a set of domain specific ``middle'' or ``low level'' simulators to be used only when a fine grained level of detail is necessary (e.g.,~OMNeT++~\cite{omnet}, ns-3~\cite{ns3}, SUMO~\cite{sumo}). The switch between coarse and fine grained models can be automatic or triggered by the simulation modeler. For example, if a given simulated area is populated by too many wireless devices, then a detailed simulation model could assess network capacity or congestion problems. The main issues with this multi-level approach are the interoperability among the simulators and the design of the inter-model interactions (e.g.,~synchronization and state exchanges at runtime between model components).

\begin{figure*}[ht]
\centering
\begin{subfigure}
\centering
\includegraphics[width=.7\linewidth]{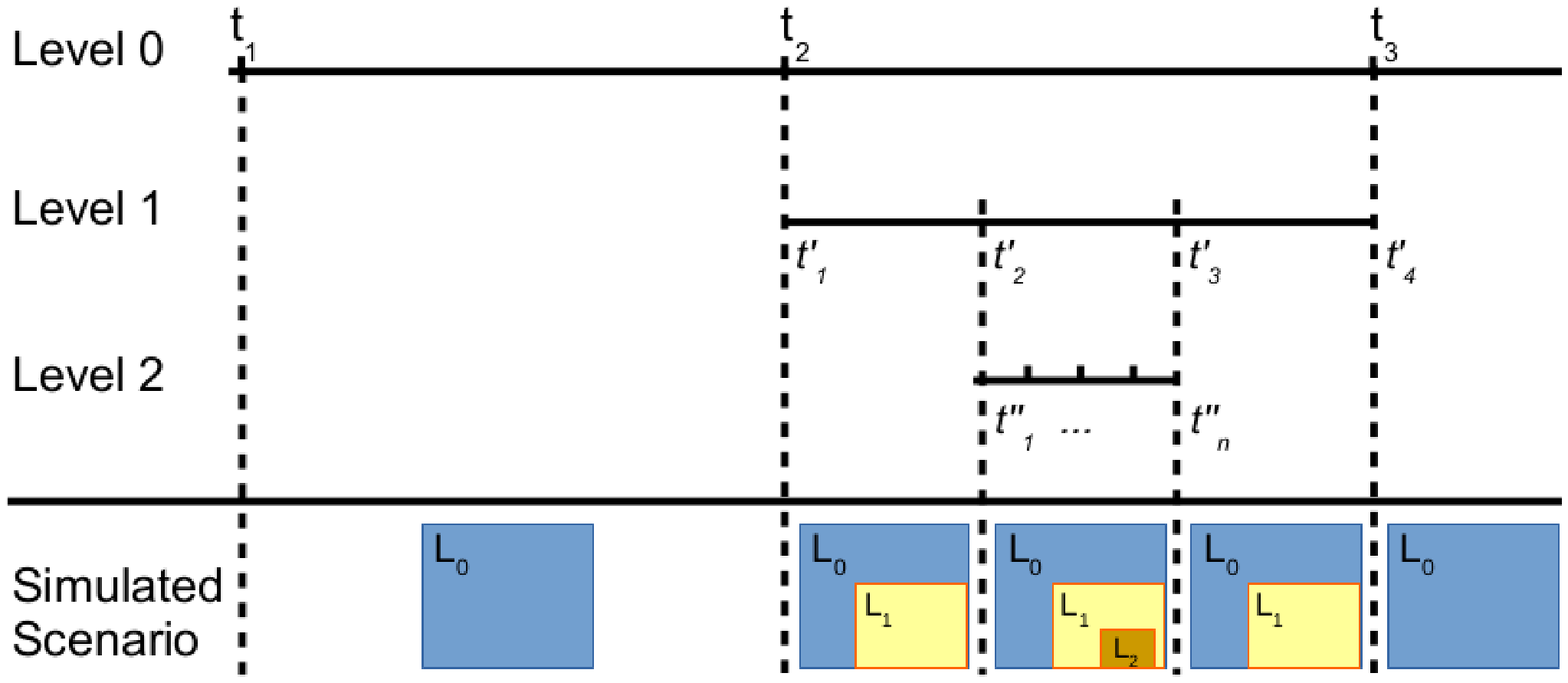}
\caption{stratified multilevel simulation.}
  \label{fig:multilevel}
\end{subfigure}%

\begin{subfigure}
\centering
\includegraphics[width=.7\linewidth]{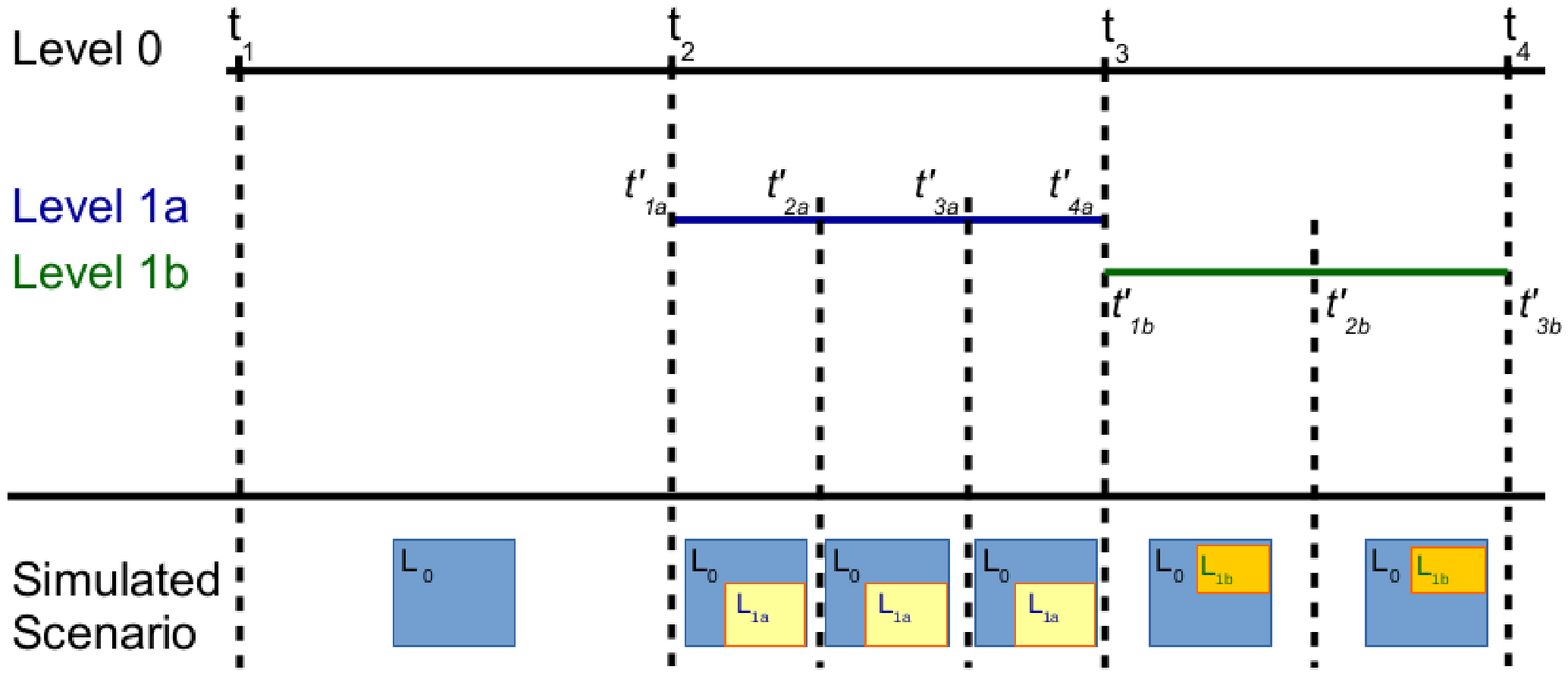}
\caption{separate simulation models.}
  \label{fig:separate}
\end{subfigure}%
\caption{Hybrid simulation: examples of stratified multilevel simulation and separate simulation models.}
\label{fig:hybrid-simulation}
\end{figure*}

Figure~\ref{fig:hybrid-simulation} shows two examples of hybrid simulation scenarios. In both examples, at bootstrap the whole scenario is executed at Level 0 ($L_0$, that is, with a minimal details). 
Hence, the high level simulator manages the evolution of all the model components and their interactions following a time-stepped synchronization approach~\cite{gda-mswim-2004}. 
In the example shown in Figure \ref{fig:multilevel}, at timestep $t_2$, it is found that a part of the simulated scenario (for example a specific zone in the simulated area or a specific group of modeled nodes) has to be simulated with more details. This means that, in the figure, a part of the simulated area is still modeled at Level 0 while a specific zone is now managed by the Level 1 model. 
If necessary, in the following of the simulation, a specific area can be further detailed using a Level 2 simulation model. To simplify this discussion, if we consider only two levels, then all the model components managed by the Level 0 simulator are evolved using $t$-sized timesteps and all the others use $t'$-sized timesteps. Timestep $t_2$ (that is the same as $t'_1$ for Level 1) is the moment in which a part of the model components is transferred from the coarse grained simulator to the finer one. The components at Level 0 will jump from $t_2$ to $t_3$ while the components simulated at Level 1 will be updated at $t'_2$, $t'_3$ and $t'_4$ (that is the same as $t_3$ for Level 0). Following that, since there is no more need for the fine-grained level of detail, all the components simulated at Level 1 are transferred again to the Level 0 simulator. Following the constraints imposed by the time-stepped synchronization algorithm, all the interactions among Level 0 simulated components can happen at every coarse grained timestep while the interactions at Level 1 can happen at every fine grained timestep. Finally, the interaction between components managed at different levels can happen only at the coarse grained timesteps. That is, when there is a match between the timesteps at the different levels.

This was an example of a stratified generation of multiple levels of simulation models. Figure \ref{fig:separate} shows a different example, according to which different simulators are separately employed to provide finer-grained simulations in different areas of the simulated world. 
For the sake of simplicity, in the figure the two Levels 1a and 1b have been triggered during different simulation time intervals, i.e., Level 1a occurs during the $[t_2, t_3]$ Level 0 simulation time interval, while Level 1b occurs during the $[t_3, t_4]$ Level 0 simulation time interval. 
In this example, it is important to notice that the two finer level simulators might be executed concurrently if they are executed in different regions of the simulated area; otherwise, when the two simulators act in the same area, they must be somehow properly synchronized.
In this case, since in the example these finer grained simulation models (Level 1a and Level 1b in the figure) are independent, the timesteps employed by these two simulators might be of different (time) size. (However, the constraint, that a single timestep of the higher level simulator L0 should be composed of multiple timesteps of a low level simulator, remains.) 

It is worth noticing that Figure \ref{fig:hybrid-simulation} emphasizes the differences on the granularity of the temporal dimension related to different levels of simulation. However, there are other aspects to deal with when crossing these levels, e.g. objects, fidelity, resolution, etc. 
In fact, when multiple simulators and/or different simulation methodologies are utilized, the transfer of objects and data, related to simulation entities, can result as a hard task. There might be problems concerned with model fidelity, granularity, as well as interoperability among different simulators and models.
The level of granularity of each of these aspects is related to the specific simulation that has to be performed. For instance, in some cases one might need increasing the spatial resolution (since a higher precision is required on the position of the simulated entity), while in other ones it might be necessary adding novel features to be considered (e.g.,~imagine a simulated entity representing a user, equipped with a mobile device, entering an area with an available WiFi networking communication infrastructure; thus, wireless networking protocols might have to be simulated).
All these simulation aspects and features have to be considered in order to be interoperable for all these different levels of simulation.

The use of a hybrid simulation approach does not change the total number of simulated nodes active in the whole simulation model, but the level of detail used in the simulation is adapted to the needs of the simulation, at runtime. 
In other words, the level of detail, and the specific features that are simulated at a finer (coarser) level of detail, is adapted and changed during the simulation, dynamically. Hence, it is possible to obtain a better scalability with respect to traditional simulation (monolithic or PADS) approaches. On the other side, it is clear that hybrid modeling (as every kind of model approximation) might introduce some (approximation) error in the analysis. As in every simulation, appropriate verification and validation techniques need to be used~\cite{Eldabi:2016:HSH:3042094.3042274}.

Another aspect that must be considered is usability. It is well known that PADS technologies, after many years of research and development, still have usability issues~\cite{Fujimoto:2016:RCP:2892241.2866577,gda-hpcs-11}. The hybrid simulations can have the same usability problems combined to the interoperability issues of using multiple simulators. Even if, the agent-based approach can reduce some of the PADS problems and the usage of domain specific simulators can simplify the modeling of complex systems, there are still problems that need to be addressed. In our view, the main ones are: 1) the need of methodologies and tools for the definition of conceptual models; 2) the availability of simulation paradigms that are easy to use; 3) well-defined interfaces for the interoperability among simulations (e.g.~simulation standards that are widely accepted and used); 4) new approaches for the composability of simulators; 5) automatic mechanisms for the deployment and the management of simulators on distributed (e.g.~cloud) infrastructures. It is clear that, most of these problems are not specific of the hybrid simulation approach but more general.

%% file: sec_casestudy.tex
\section{Case Study}
\label{sec:casestudy}

\subsection{Smart territories}
As an application scenario, we consider a main use case concerned with the need to provide smart services to territories, being cities or more decentralized areas. In particular, we focus on ``smart shires'', a novel view of decentralized geographical spaces able to manage resources (natural, human, equipment, buildings and infrastructure) in a way that is sustainable and not harmful to the environment~\cite{gda-simpat-2017,smartshires,smartshires_abps}. The idea is to create novel, smart and cheap services, easily deployable without the need of costly infrastructures, that would improve the life of citizens and tourists.

The need for cheap solutions forces the use of crowd-sensed and crowd-sourced data coming from the IoT. Sensors are relatively cheap in terms of costs. Thus, their deployment in a countryside is feasible. These sensors need to be interconnected through the use of smart communication approaches~\cite{Petrolo:2014}. Data sensed by the sensors' devices are ma\-na\-ged by a distributed information processing system, hence enabling a context-aware data distribution~\cite{Bellavista:2012}.

A wide range of application scenarios are possible, ranging from proximity-based applications (e.g.~proximity-based social networking, advertisements for by-passers, smart communication between vehicles, etc.), security and public safety support, services related to the production chain in rural environments (smart agriculture, smart animal farming) and smart traffic management systems.

\subsection {A use case with a smart market}
As a specific use-case example, recently the ``km 0'' phenomenon gained a lot of interests in Italian and European foodie circles. This abbreviation for ``zero kilometers'' signifies local, low impact primary food ingredients. The idea is to prioritize the use of local and seasonal foods, avoid the use of genetically modified organisms so as to improve the quality of provided products and promote sustainable cooking. In spite of the growing interest in local products, there are relatively few places where one can buy these products directly from the producer. Thus, customers have to look for specialized weekend farmers markets or for farm direct purchases. Customers might be single users, ethical purchasing groups or restaurant owners. And quite often, this products research reveals to be not a simple task for customers. Thus, smarter scenarios are possible.

Let's imagine a service that allows consumers subscribing to the availability of a certain product. Upon availability of such a product by a producer (e.g.~the farmer), he can publish a notification, which informs subscribers of product availability plus other related information such as, his presence in next, near markets or other possible purchasing opportunities.
In view of such details, the consumer can plan to visit the market (so as to have the opportunity to select the products directly), order some specific items, quantities and so on. There are plenty of publish-subscribe mechanisms that might help these producer/consumer interactions in order to build smarter services. 
However, more sophisticated services are possible. 
The arrival at the market might be guided by services that help the user find a parking spot. The location might not be equipped with associated parking, thus the service might reduce the time spent by customers to find proper parking, hence improving the quality of experience, increase their willingness to come back in the future, and also reducing vehicle emissions.

Once there, the market could be crowded with several (apparently similar) producers and the customer might not know the exact location of producers. Thus he might need to be guided to the producers locations dynamically. Moreover, he might also be interested in finding other possible interesting products.

To cope with these issues, producers can provide information on the fly, thanks to proximity-based services that may guide customers in a smart and effective market tour.
Based on the available technologies of the market, such services can be deployed in different ways. For instance, if a wireless infrastructure is available, then all the communications can pass through this network. Otherwise, some ad-hoc solution should be dynamically built, with producers that exploit their smart devices (e.g.~smartphones) to build multihop wireless communication and information dissemination strategies~\cite{smartshires_abps}.
Moreover, in case of intermittent connections, seamless communication strategies should be employed using multihoming~\cite{Ferretti2016390}. Message dissemination about the market (e.g.~advertisements, general information) might be viably performed using some kind of epidemic dissemination protocol over a dynamic, opportunistic ad-hoc overlay, used in conjunction with application filtering techniques~\cite{simplex,Ferretti2013481,Wirtz:2014}.

\subsection{The distributed hybrid simulator}
\begin{figure*}[ht]
\centering
\includegraphics[width=12cm]{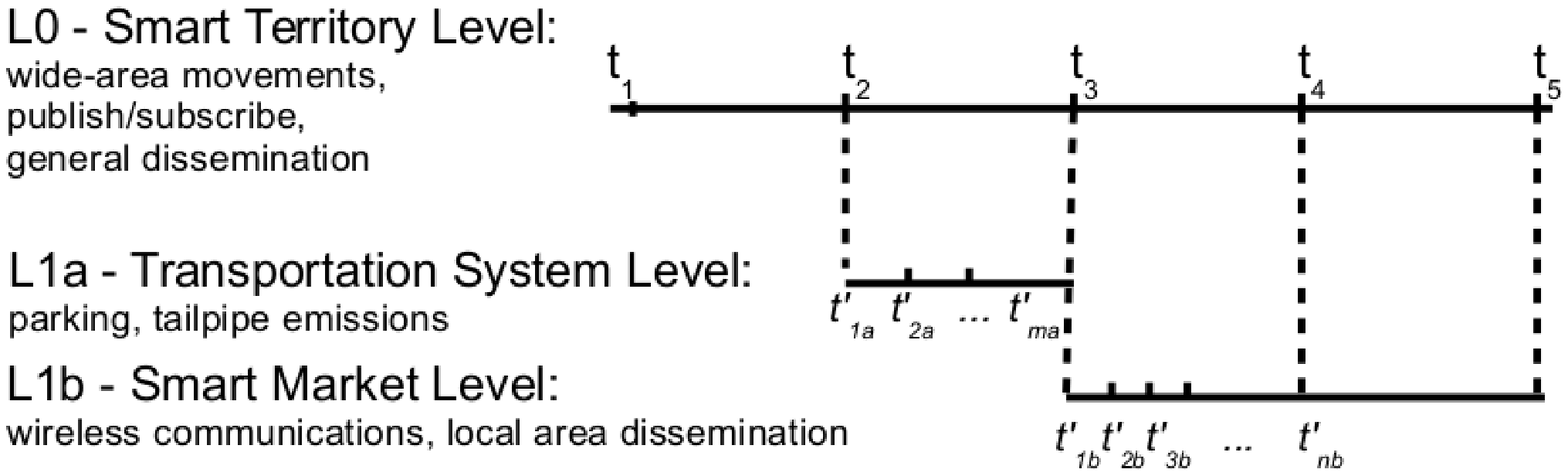}
\caption{Smart Territory/Market hybrid simulation.}
\label{fig:usecase-multilevel}
\end{figure*}

The efficient simulation of such a wide scenario in a smart territory is not an easy task, since it involves several activities concerned with different domains and requires very different levels of granularity. Needless to say, a solution might be to employ a classic agent-based simulator to model all the aspects concerned with this use case. However, this would require implementing all the features and issues that need to be considered in this scenario. Moreover, as already discussed the use of a single simulator, that keeps the whole simulation at a high level of detail, might not scale.
Indeed, in this case, hybrid simulation can come into the picture. One can imagine different levels of granularity, as shown in Figure \ref{fig:usecase-multilevel}. We employ three different simulation models, that act at two different levels, i.e., a coarse simulator that can trigger the execution of two Level 1 simulators that can be executed sequentially. Such simulators are distributed and executed in different hosts. The coarse level ($L_0$ in the figure) simulates the whole smart territory, where different actors produce products, subscribe their interests and move towards different geographical areas. This has been implemented using an agent-based simulator with adaptive PADS capabilities~\cite{gda-simpat-2017-part}.

The arrival of customers (through a transportation system) to the location, the parking deployment strategies in the neighborhood of the market, etc.~can be simulated using some specific tool able to model transportation systems and related issues, such as the analysis of tailpipe emissions (see $L_{1a}$ in Figure  \ref{fig:usecase-multilevel}).
This enables studying the goodness of the parking and transportation solutions, and estimating the amount of time users spend entering the market, if they will.
This level has been implemented by resorting to the MATLAB based ADVISOR tool~\cite{advisor}. 

Finally, once pedestrian users enter the smart market (or the location where it takes place)
there is the need to simulate the specific interactions within that specific area. In this case, more simulation details (and probably a different simulator) are needed to consider wireless communication issues, fine-grained interactions and movements. Thus, a more detailed level of simulation (based on a domain specific simulator) is triggered (i.e.~$L_{1b}$ in the figure). In this case, each simulation step of the coarse grained simulation layer (e.g., $t_3, t_4$ of $L_0$ in Figure~\ref{fig:usecase-multilevel}) is decomposed into multiple substeps at the fine grained layer. Following this approach, the Level 1 simulator is able to notify Level 0 with its simulation advancements.
The specific interactions within the smart market impose more simulation details, so as to consider wireless communication issues, fine-grained interactions and movements. 
Thus, an instance of OMNeT++ simulation has been implemented.

The main issue is to provide means to let the simulators interact. In fact, when needed $L_0$ triggers 
the execution of $L_{1a}$, passing some arguments serving as configuration and initialization parameters. Then, the finer level simulator must run for a certain amount of timesteps, producing outputs that are passed to $L_{1b}$.
Results are passed to $L_0$ at the end of each timestep. In turn, Level 0 must ask simulators at Level 1 to continue or end their simulation.
In the following, we outline the main characteristics of the three simulators and then we show how they are managed and coordinated into the distributed hybrid simulation system.

In terms of the simulation model implemented by the hybrid simulator that is proposed in this paper, a rigorous and correct modeling approach would be to define a conceptual model, following a methodology such as that described in \cite{Robinson2008} and \cite{Robinson2008-2}. However, in this work we prefer to focus on the implementation aspects resulting from the combination of PADS, monolithic DES and continuous simulation tools. In other words, in this experience we investigate the possibility of building efficient distributed hybrid simulators and initiate a discussion on the performance metrics that should be used for evaluating this kind of simulators.

\subsection{Level 0: agent-based simulator}
\label{sec:l0}
Smart Shire Simulator ($S^3$) is a prototype based on the GAIA/ART\`IS simulation middleware~\cite{pads,gda-simpat-2014}. ART\`IS permits the seamless sequential/parallel/distributed execution of large scale simulation runs using different communication approaches (e.g.~shared memory, TCP/IP, MPI) and synchronization methods (e.g.~time-stepped, conservative, optimistic). The GAIA part of the software tool aims to ease the development of simulation models with high level application program interfaces. Furthermore, it implements communication and computational load-balancing strategies (based on the adaptive partitioning of the simulation model), for reducing the simulation execution time.

The current version of $S^3$ implements a limited set of functionalities. The many elements composing the smart shire are represented as a set of interacting entities. Some entities are static (e.g.~sensors, traffic lights and road signs) while the others (e.g.~cars and smart-phones) follow specific mobility models. All the entities in the simulated model are equipped with a wireless device. 

The interaction among entities is based on a ``geo-location Priority-based Broadcast'' (geoPbB) strategy that implements a novel probabilistic broadcast approach with geolocation filtering capabilities. In geoPbB, every message that is generated by a node is broadcast to all the nodes that are in proximity of the sender. The message contains a Time-To-Live (TTL) and geographical information to limit its lifespan and the forwarding is based on three conditions. 
First, the relay of a message is performed only with a certain probability (i.e.~probabilistic broadcast). Second, the distance between the sender and receiver is measured before relaying a message; in substance, a message is forwarded only if the distance between the nodes is larger than a given threshold.
Put in other words, only nodes in the external annular ring of the transmission range participate to the gossip procedure.
Under the implementation viewpoint, this can be done using a positioning system (e.g.~GPS) if available. Otherwise, the network signal level associated to each received message is used. Third, the message is relayed only within a certain geographical area (geolocation filtering). In fact, we assume that the geographical position of the message originator is an important data in a proximity based service, such as those we are dealing with in this case study. Hence, once a message ``leaves'' a specific area of interest, the message is discarded and not relayed anymore.

\subsection{Level 1a: ADVISOR based simulator}
\label{sec:other}
The transportation system related issues are implemented through a simulator based on the ADvanced VehIcle SimulatOR (ADVISOR)~\cite{advisor}.
ADVISOR is a MATLAB/Simulink based simulation tool for the analysis of the performance and fuel economy of conventional (gasoline/diesel), electric, and hybrid vehicles. 
It allows interchanging a variety of components, vehicle configurations, and control strategies.
The goal of the simulator is to allow testing efficiency of automobiles, especially in terms of tailpipe emissions, fuel economy, acceleration and grade sustainability.
To this aim, the simulator works using a component-based approach, where components are typically modeled through a set of equations and quasi-steady approximations.
While the typical use of the tool is based on a graphical interface, it provides means to perform batch simulations. We employed the batch execution mode to make the simulation work and interact with other components of our hybrid simulation software.

Thanks to features provided by the ADVISOR simulator, Level 1a simulates vehicles arriving at the neighborhood of the market place and the parking activity of these customers. Level 1a measures the amount of emissions, that are then passed to the Level 0 simulator.
Moreover, this level of simulation has detailed information on vehicles arriving to the car park; thus, this simulator establishes how many users park their vehicle and enter the market place.
In substance, Level 1a provides Level 1b with the number of novel customers entering the market place. Thus, the execution of these two Level 1 simulators is sequential.

\subsection{Level 1b: OMNeT++ simulator}
\label{sec:omnet}
The fine grained simulator of the smart market was implemented using OMNeT++ v.~4.4.1, with the INET framework v.~2.3.0. It simulates a grid of fixed nodes (during the tests, a $10\times10$ grid was used), representing the market sellers. Each seller is equipped with a WiFi enabled technology. In the simulated scenario, no WiFi infrastructure was present, hence nodes organize themselves as a MANET exploiting DYMOUM~\cite{www_dymo-um}, an implementation of the Dynamic MANET On-demand (DYMO) routing protocol~\cite{ietf-manet-dymo-26}.

In such a MANET a number $N$ of mobile nodes, representing pedestrian users, was introduced by the higher Level 0 simulator. These $N$ nodes are equipped with a mobile device with a WiFi networking technology. Pedestrian users move at walking speed. The user application running on the mobile client broadcasts messages looking for the identifier of the specific seller. The seller replies with his geographical position. All these messages are delivered through the mentioned MANET routing protocol. Based on the provided position, the mobile user moves towards his destination.

\subsection{Interaction among the simulators}
An interaction among the Level 0 and Level 1 simulators is needed in order to let simulators interoperate and synchronize. This means that simulators must exchange their inputs and outputs, and that $L_0$ must coordinate the execution of Level 1 simulators.
Moreover, the implementation of the different simulators impose their execution into different operating systems. In fact, $L_0$ and $L_{1b}$ are based on Linux systems, while the MATLAB based $L_{1a}$ runs on top of a Windows system. 
With this in view, the hybrid simulator is a distributed system where different instances of simulators can be executed.
A message-passing approach is utilized, which has been realized through the use of TCP socket connections.

\begin{figure*}[ht]
\centering
\includegraphics[width=12cm]{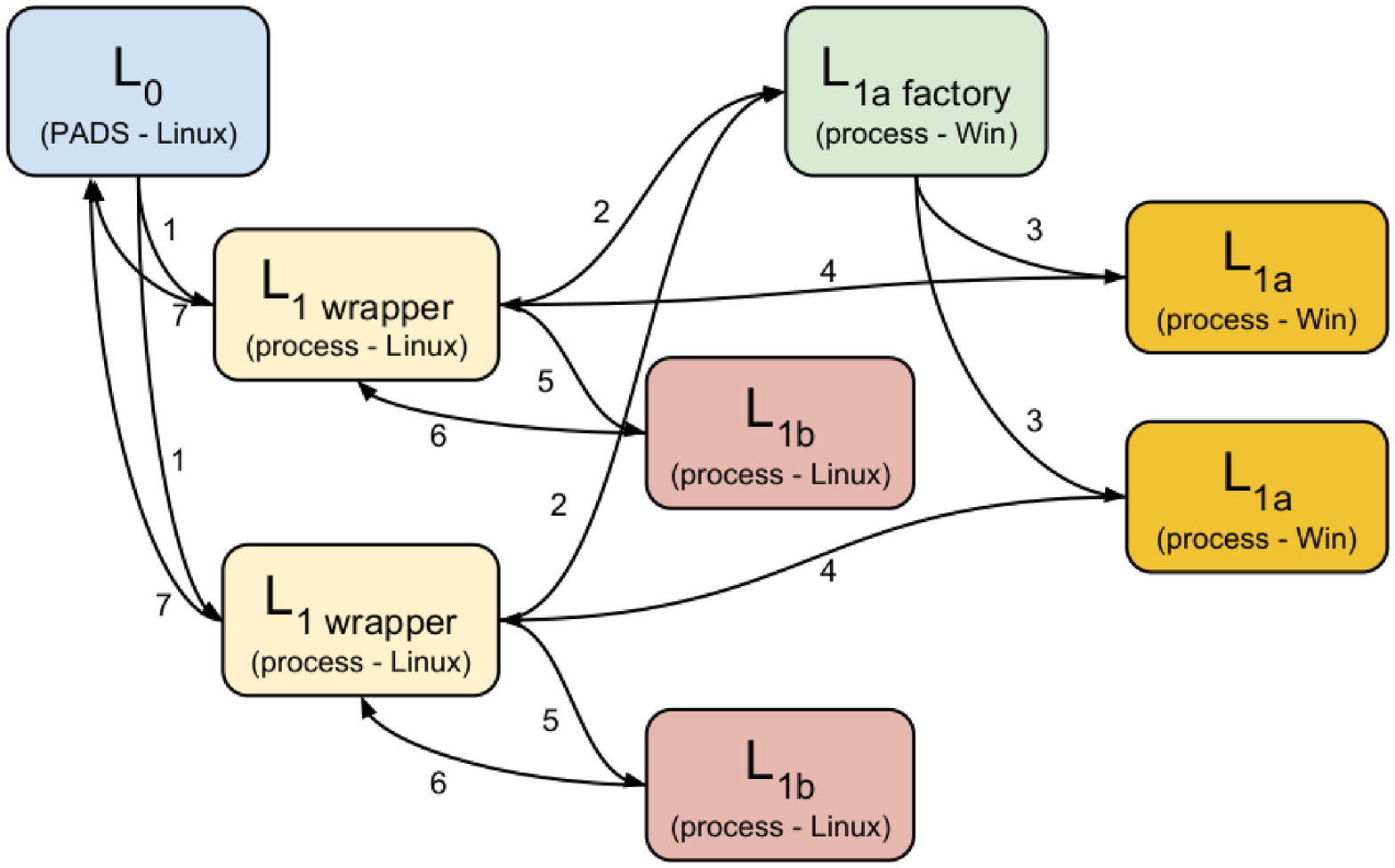}
\caption{Distributed hybrid simulation architecture (with two depicted Level 1 simulation instances).}
\label{fig:archi}
\end{figure*}

Figure \ref{fig:archi} shows the distributed architecture of the hybrid simulator. 
The $L_0$ instance can run as a PADS into one or multiple Linux hosts. 
Since the two Level 1 simulators are thought to be executed sequentially, for the sake of a simpler implementation, we created a Level 1 wrapper that coordinates the execution of the two Level 1 simulators and synchronizes their sequential activities with the Level 0 simulator. In the figure, as an example, we depicted two instances of such a wrapper ($L_{1\textnormal{ wrapper}}$), mimicking that two instances of Level 1 have been generated at a certain point of the simulation. 
Thus, $L_0$ triggered two $L_1$ wrappers, passing all the configuration details (arrows (1) in the figure).

The $L_{1\textnormal{ wrapper}}$ has been realized as an OMNeT++ component. It sequentially launches an instance of $L_{1a}$ and, once completed, the OMNeT++ $L_{1b}$ instance, passing the output of $L_{1a}$ to $L_{1b}$, i.e.,~the amount of novel customers entering the market.
In order to do it, since $L_{1a}$ instances run on Windows operating systems, we created a Windows application that basically acts as a listening server waiting for novel connections ($L_{1a\textnormal{ factory}}$ in the figure). Thus, $L_{1\textnormal{ wrapper}}$ connects to $L_{1a\textnormal{ factory}}$ passing all the configuration parameters (arrows (2) in the figure). Then, $L_{1a\textnormal{ factory}}$ generates an instance of a properly configured $L_{1a}$ instance (arrows (3) in the figure).
Once completed, $L_{1a}$ returns its output to $L_{1\textnormal{ wrapper}}$ (arrows (4) in the figure), that in turn generates a novel instance of the $L_{1b}$ OMNeT++ simulation (arrows (5) in the figure). Once completed, the $L_{1b}$ OMNeT++ simulator returns its output to $L_{1\textnormal{ wrapper}}$ (arrows (6) in the figure), that in turn returns its output to L0 (arrows (7) in the figure).

While these Level 1 instances are active, $L_0$ waits for messages from the $L_{1\textnormal{ wrappers}}$ at properly configured socket connections. At the end of each Level 0 timestep, $L_{1\textnormal{ wrapper}}$ sends a set of messages which describe its status and waits for a response. $L_0$ receives the data sent by $L_{1\textnormal{ wrapper}}$ and decides what message has to be sent (i.e., continue or end the lower level simulation). The TCP connection is maintained until the $L_0$ simulator decides that the lower level simulation must end.

In essence, this strategy allows interactions between the simulators without requiring a complete re-engineering of the simulators. In this approach, the higher level simulator must be able to freeze the simulation of certain parts of the simulated scenario, waiting for updates from other sources. Moreover, lower level simulators should be enabled to obtain input from outside, and notify results outside. However, no knowledge on the external simulators are needed. This is an example demonstrating that existing products can be employed to create more complex multi-level simulations.

\subsection{Implementation details}
The $L_{1a\textnormal{ factory}}$ process, and its child processes, work on a Linux-like Cygwin environment that operates on top of a Windows 8.1 operating system. The orchestration between the $L_{1\textnormal{ wrapper}}$ and the $L_{1a}$ subsystem is a little bit complex because each $L_{1a}$ process, as shown in the Figure~\ref{fig:archi}, is composed of a couple of processes, a father and its child. The father inherits, from the $L_{1a\textnormal{ factory}}$ process, the TCP connection with the $L_{1\textnormal{ wrapper}}$ process from which receives the simulation parameters. The father increments a counter and uses this unique value to generate the name of a couple of files that uses as input and output with its process child. The father writes the simulation parameters into the input file and creates a child process giving to it the pathnames of both the input and output files. Then, the father waits for the termination of the child process. The child process runs a bash script that executes the MATLAB simulation launching a MATLAB script into a MATLAB engine. This is done by means of the following command line:

\texttt{
\\
\indent matlab -nodesktop -nosplash -r "L1a\_simfunc(\$\{fileinput\},\\ 
\indent\indent\indent\indent\indent\indent\indent 	\$\{fileoutput\}); quit force"
\\
}

The two initial parameters select the console version user interface of the MATLAB engine and avoid its graphical user interface. The rest of the command line provides the MATLAB engine with the name of the function ($L1a\_simfunc$) that implements the $L_{1a}$ simulations and the name of the input and output files to be used. The final command ($quit\ force$) closes the console of the MATLAB engine at the end of the simulations. The simulation function $L1a\_simfunc$ uses the parameters loaded from the input files and executes a sequence of MATLAB functions that sets the ADVISOR scenario and simulates the transportation subsystem. Finally, the MATLAB script saves the simulation results into the output file, closes the MATLAB engine and terminates the child process. At the end of the MATLAB script, the bash script terminates the child process and signals the father process. The father process wakes up and loads the simulation results from the output file, then it sends the results through the TCP connections to its $L_{1\textnormal{ wrapper}}$. Finally, the father deletes both the input and output files and terminates itself and its TCP connection.

%% file: sec_perf.tex
\section{Performance Evaluation}
\label{sec:perf}

Assessing the performance of a hybrid simulator can be a hard task, since several components must be evaluated as well as their interactions. When these components are distributed, the network can be a further variable that can strongly influence the performance of the simulation. With this in view, in this section, firstly we evaluate the scalability of Level 0 simulator alone.  Then, since Level 1a ($L_{1a}$) is ancillary to Level 1b ($L_{1b}$), we assess $L_{1b}$ and its performance with the Level 0 ($L_0$) simulator. Finally, we assess the performance of the whole distributed hybrid simulator (that is $L_0$ + $L_{1a}$ + $L_{1b}$).

All the results reported in this section are averages of multiple independent runs. This performance evaluation has been performed using two hosts interconnected by a Fast Ethernet LAN. The Linux components ($L_0$ and $L_{1b}$ simulators) were run on a DELL R620 with 2 CPUs and 128 GB of RAM. Each CPU is a Xeon E-2640v2, 2 GHz, 8 physical cores. Each CPU core supports Hyper-Threading and therefore the number of logical cores is 32. The computer is equipped with Ubuntu 14.04.5 LTS, GAIA/ART\`IS version 2.1.0, OMNeT++ v.~4.4.1 (with the INET framework v.~2.3.0). The $L_0$ simulator $S^3$, the $L_{1\textnormal{ wrapper}}$, $L_{1\textnormal{ factory}}$ and $L_{1b}$ OMNeT++ models used for the multi-level simulator will be freely available as source code in the next release of GAIA/ART\`IS~\cite{pads} or upon request.
The Windows-based components $L_{1a\textnormal{ factory}}$ were executed on a host with Windows 8.1 Pro operating system, Intel core i7-5600U, 2.5 GHz, 8 GB of RAM.

To sum up, in this performance evaluation the whole hybrid simulation was a distributed simulation, embodying a parallel simulation at Level 0 and multiple sequential simulations at Level 1.

\subsection{Level 0: agent-based simulator}
The performance evaluation of $S^3$ is based on a bidimensional toroidal space (with no obstacles) that is populated by a given number of devices called Simulated Entities (SEs) (that are static and dynamic nodes in the smart shire such as cars, people, signaling devices and so on). A subset of the SEs follows a Random Waypoint (RWP)~\cite{rwp} mobility model while the others are static. The interaction among SEs is based on proximity and implements the geoPbB strategy previously described in Section~\ref{sec:l0}. In particular, each node is enabled with a caching mechanism to discard (some of) the duplicated messages (based on a last-recently-used replacement algorithm) and with a geolocation filtering mechanism. Moreover, with the aim of reducing the amount of messages that are relayed by nodes, we implemented an additional filtering scheme, according to which nodes, whose distance from the sender is lower than a certain threshold, do not relay the message. Hence, only nodes in the external annular ring of the transmission range participate in the gossip procedure.

The message filtering strategy described above is integrated in the gossip dissemination mechanism as defined in our smart shire proposal. For this reason, this model-specific filtering strategy is implemented at the Level 0 simulation model. It is well known that the overhead caused by the model level communication has a big impact on the scalability of PADS. For this reason, many relevance filtering techniques have been developed for being used in both Distributed Interactive Simulation (DIS) and High Level Architecture (HLA) standards~\cite{Bassiouni:1998:RFD,Bassiouni:1997:PRA:259207.259209}. These techniques are more general than the gossip filtering approach implemented in $S^3$ and their integration in GAIA/ART\`IS is left as future work.

Table~\ref{table:model} shows the main parameters of this performance evaluation. Some of such parameters are strictly dependent on the specific scenario characterized by the geographical and architectural issues of the smart shire deployment, while others are more general. In our view, simulation tools are necessary to support the design of IoT architectures and for the tuning of the many runtime parameters.

\begin{table*}[h]
\begin{center}
	\begin{tabular}{ | l | p{7cm} |}
	\hline
  \textbf{Model parameter} & \textbf{Description / Value} \\ \hline
	Number of $SEs$ & [1000, 32000] \\ \hline
  Mobility of $SEs$ & 50\% Random Waypoint (RWP)\newline 50\% static  \\ \hline
  Speed of RWP & Uniform in the range [1,14] spaceunits/timestep\\ \hline
  Sleep time of RWP & 0 (disabled)\\ \hline
  Interaction range & 250 spaceunits\\ \hline
  Density of $SEs$ & 1 node every 10000 $\textnormal{spaceunits}^2$\\ \hline
  Forwarding range & $>225$ spaceunits\\ \hline  
  Simulated time & 900 timesteps \\ \hline
  Simulation granularity & 1 timestep = 1 timeunit \\ \hline
  Time-To-Live (TTL) & 6 hops \\ \hline
  Dissemination probability (gossip) & 0.2 \\ \hline
  Prob. of each SE to generate a new message & 0.001 (per timestep)\\ \hline
  Cache size (positions) &  128 \\ \hline
  Geofiltering distance &  1000 spaceunits \\ \hline
  Max forwarded messages per SE &  10 (per timestep) \\ \hline
  \end{tabular}
\end{center}
\caption{Simulation model parameters.}
\label{table:model}
\end{table*}

\begin{figure}[h]
\centering
\includegraphics[width=8.5cm]{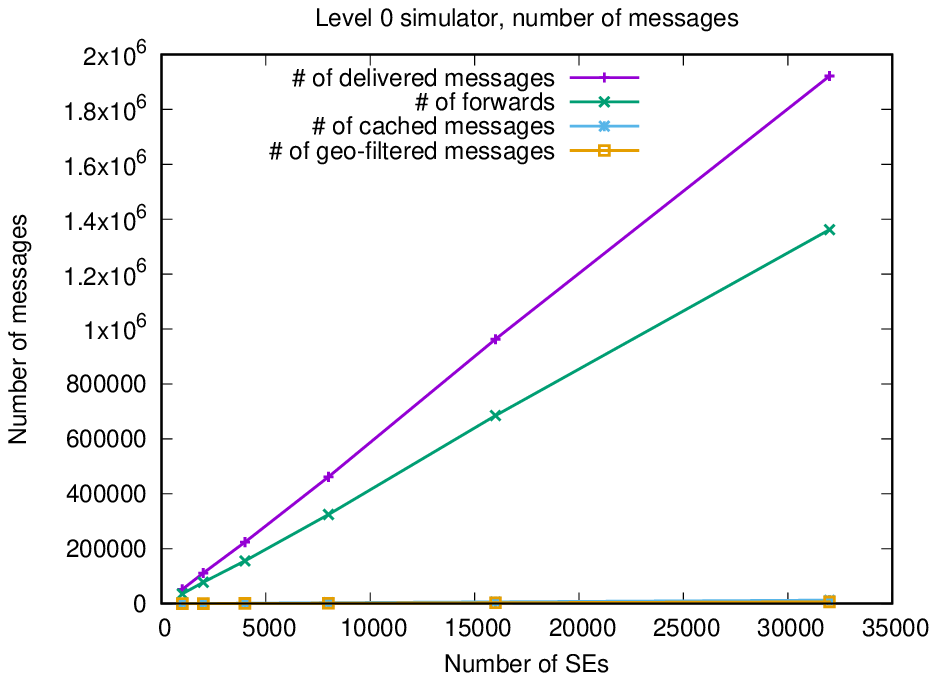}
\caption{Model characterization: number of simulated messages, exchanged by SEs during the simulations.}
\label{fig_messages}
\end{figure}

\begin{figure}[h]
\centering
\includegraphics[width=8.5cm]{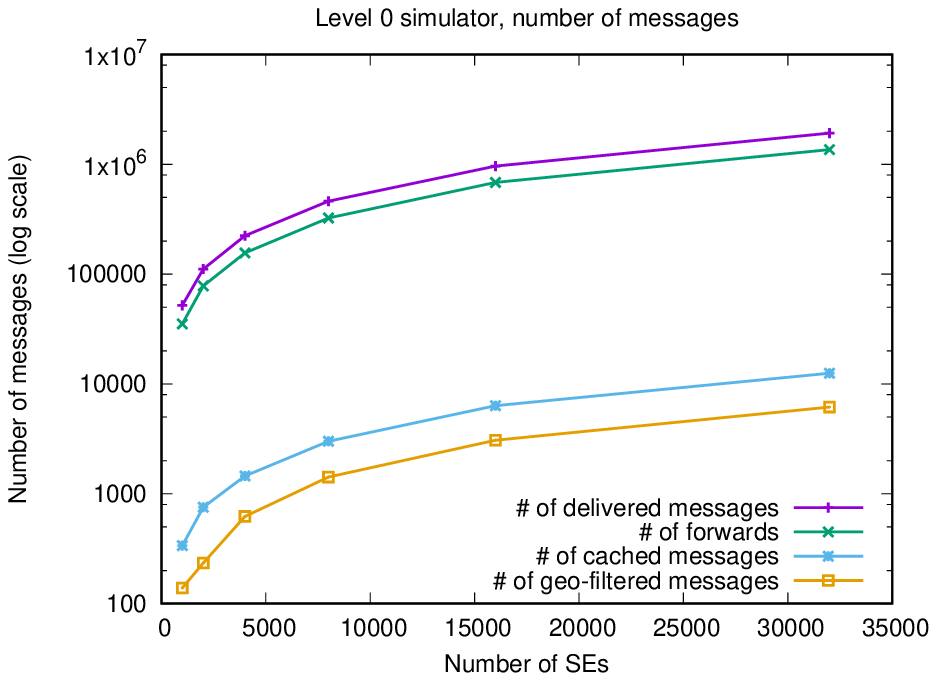}
\caption{Model characterization: number of simulated messages, exchanged by SEs during the simulations (log scale).}
\label{fig_messages-log}
\end{figure}

\begin{figure}[h]
\centering
\includegraphics[width=8.5cm]{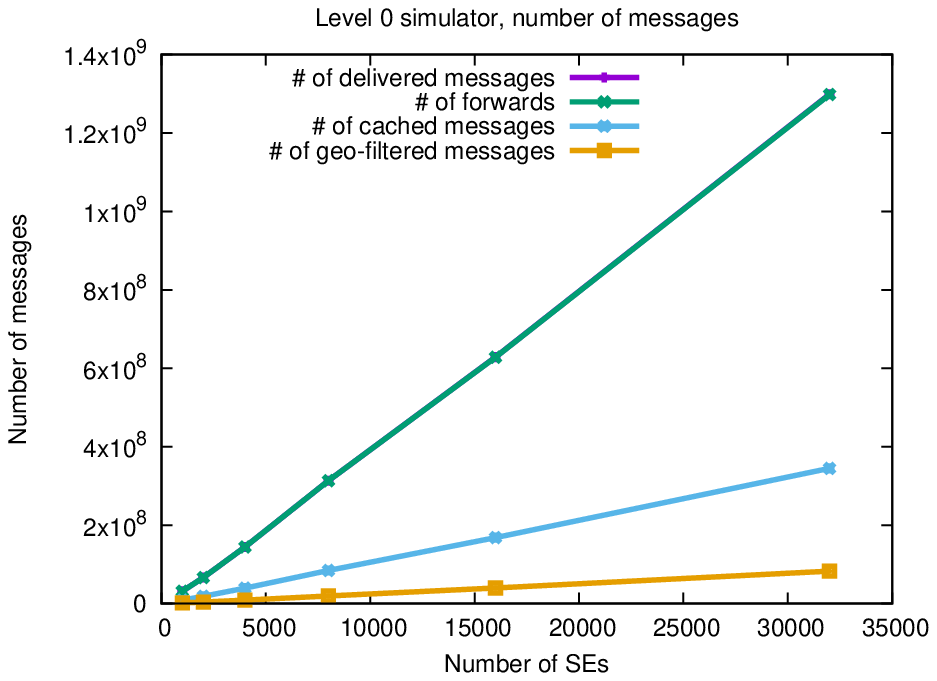}
\caption{Model characterization: number of simulated messages, exchanged by SEs during the simulations. Bad tuning of the dissemination parameters.}
\label{fig_messages-rev1}
\end{figure}

First of all, we investigate the smart shire simulation model. In particular, the geoPbB dissemination strategy is evaluated, in presence of an increasing number of SEs. Figure~\ref{fig_messages} shows the number of simulated messages, exchanged by SEs (i.e.,~simulated devices and things in the smart territory), during the simulations. In the chart, the amount of messages depends on the varied number of SEs. As shown in the figure, the number of delivered messages is high and it increases linearly with the number of nodes in the system. As frequently happens in gossip-based dissemination schemes, a large part of the delivered messages are forwards (that is, relayed messages). The impact of both caching and geofiltering mechanisms is almost negligible. More specifically, as shown in Figure~\ref{fig_messages-log} the caching is slightly more efficient, in terms of filtered messages, than geofiltering. The first result of this performance evaluation is two fold. A proper tuning of the dissemination parameters avoids the generation of a massive amount of delivered messages in the simulated system. This aspect is fundamental for the scalability of both the smart shire system and the simulator. On the other hand, the overhead in the geoPbB dissemination strategy is too high and the filtering mechanisms that are commonly used in such kind of system, such as caching and geofiltering, have unsatisfactory performances. This means that more complex changes to the dissemination protocols, based on gossip strategies, need to be designed, implemented and evaluated for the deployment on this kind of wireless networks. For completeness, Figure~\ref{fig_messages-rev1} reports the number of simulated messages, exchanged by SEs, in presence of a bad tuning of the dissemination parameters (i.e.,~dissemination probability set to 0.6, forwarding range > 100 spaceunits). As it can be seen comparing the results in Figures~\ref{fig_messages-rev1} and \ref{fig_messages}, the changes in the dissemination parameters have a huge impact on the number of delivered messages. In fact, in the bad tuning case the delivered messages increase of over 600 times.

%
%

\begin{figure}[h]
\centering
\includegraphics[width=8.5cm]{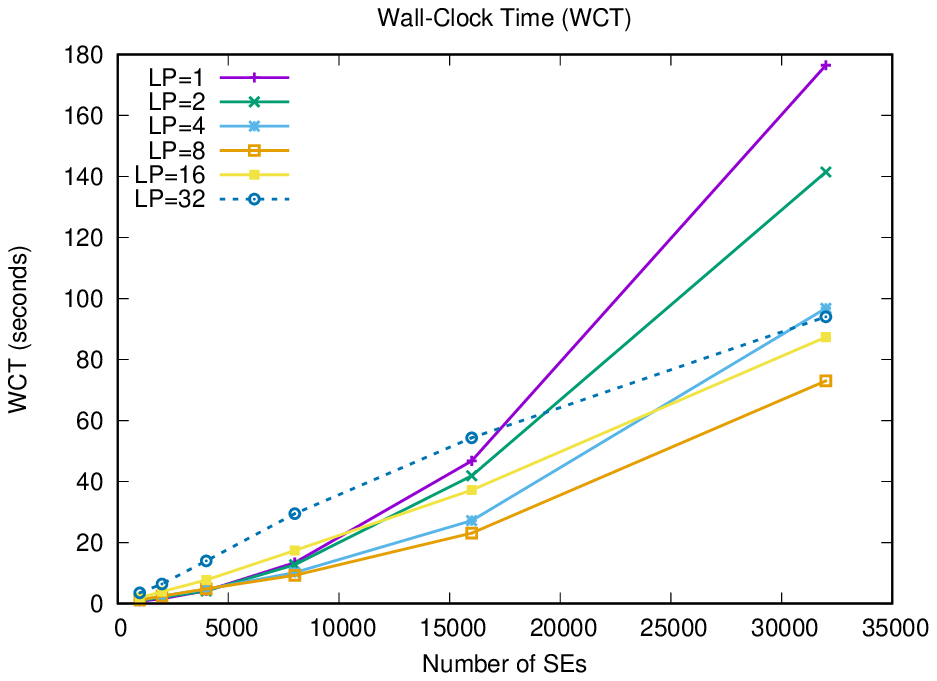}
\caption{Level 0: Scalability evaluation (WCT) -- increasing number of SEs, lower is better.}
\label{fig_L0-wct}
\end{figure}

After characterizing the message traffic of the simulated model, in the following of this section we evaluate the performance of the Level 0 simulator in both sequential (i.e.~LP=1) and parallel (i.e.~LP$>$1) execution setups (see Section~\ref{PDES_PADS}).
In the case of parallel execution, the SEs have been randomly partitioned among the LPs. In a previous work~\cite{gda-simpat-2017-part}, we have demonstrated that, in this kind of system, static partition of SEs leads to unsatisfactory results and that adaptive partitioning approaches should be preferred.
Figure~\ref{fig_L0-wct} reports the Wall-Clock-Time (WCT)\footnote{To improve the precision of WCT measuring, the execution time is calculated using the data provided by the operating system in which the simulator is executed. In presence of multi-level setups, the highest level simulator implements a blocking approach (e.g.~it waits for the completion of all the low level simulators before terminating its execution).}  required to complete a simulation run of the Level 0 simulator with respect to the number of SEs. The line for LP=1 reports the amount of time required by the simulator in a sequential (that is monolithic) setup. All the solid lines (i.e.~LP={1,2,4,8,16}) refer to setups in which only physical CPU cores are used. With LP=32 (dashed line), the number of LPs exceeds the number of physical CPU cores, in other terms, in this case all the logical CPU cores provided by Hyper-Threading are employed. When the number of SEs is very limited (i.e.~up to 2000), the sequential setup (LP=1) has the best WCT. With 4000 SEs, the best performance is obtained with 2 LPs; for higher amounts of SEs, the best setting is 8 LPs. This behavior is due to the balance between the load sharing of computation given by the parallel execution architecture and the costs of the communication between the CPU cores on which the different LPs are run. In other words, when there is a limited amount of computation, the best choice is a setup without communication among the CPU cores (this happens when LP=1). Increasing the number of SEs increases both communication and computation in the simulated model. Up to 8 LPs there is a gain in adding more computational resources, but above this number of LPs, the costs of communication is not balanced by the load sharing given by the additional CPU cores (e.g.~16 LPs). The behavior of the setup with LP=32 is slightly different, due to the characteristics of the logical cores provided by Hyper-Threading. In this case, the Hyper-Threading is unable to offer a gain. This means that, as expected, the logical cores have lower performance than the physical ones, while adding a relevant amount of costly communication.

\begin{figure}[h]
\centering
\includegraphics[width=8.5cm]{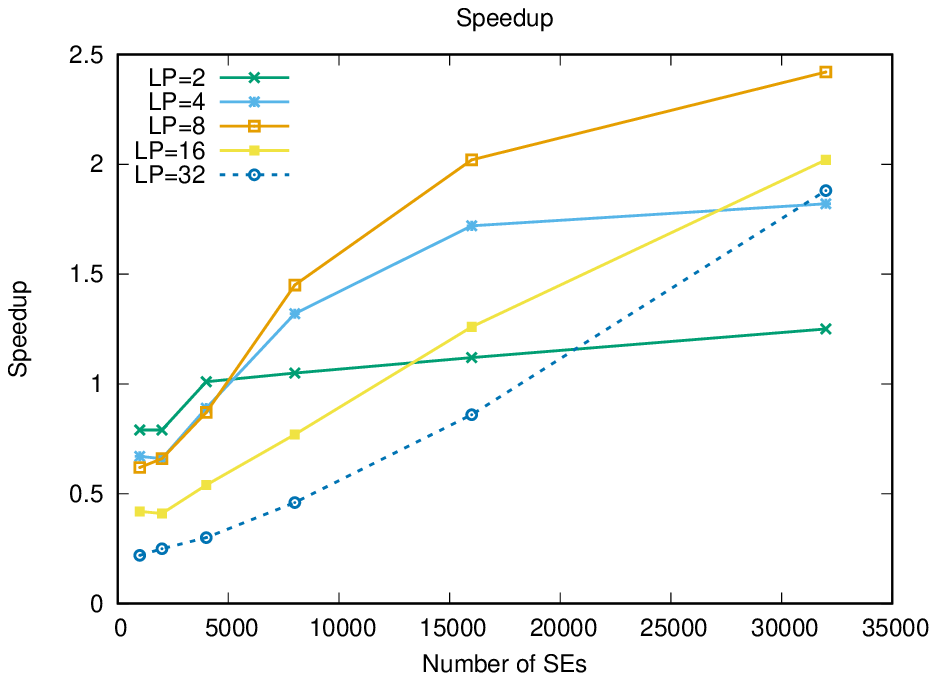}
\caption{Level 0: Scalability evaluation (speedup) -- increasing number of SEs, higher is better.}
\label{fig_L0-speedup}
\end{figure}

Figure~\ref{fig_L0-speedup} reports the speedup (i.e.~the ratio between the WCT of the sequential simulation and the WCT of the parallel execution) of the Level 0 simulator in the different setups. 
One might expect that the higher the amount of employed CPUs the higher the speedup, and that adding more CPUs to the execution architecture would lead to a linear increment of the speedup. Actually, this does not happen in most of real world applications. This is due to the fact that not all the computation can be parallelized and that the parallelization introduces some overhead (e.g.~coordination and synchronization).
The best speedup is $2.42$ and is obtained with 32000 SEs when 8 LPs are used. Given the very high communication requirements of the simulated model, it is not possible to expect large values of speedup, as those that are common in embarrassingly parallel workloads. On the other hand, as discussed above, the worst performances are obtained when low intensity computational workloads (i.e. a small number of SEs) are parallelized using many LPs (e.g. 16 or 32). In this figure, the poor performance obtained by Hyper-Threading (LP=32, dashed line) is even more evident.

\subsection{Level 0 and Level 1b}
%
%
\begin{figure}[h]
\centering
\includegraphics[width=8.5cm]{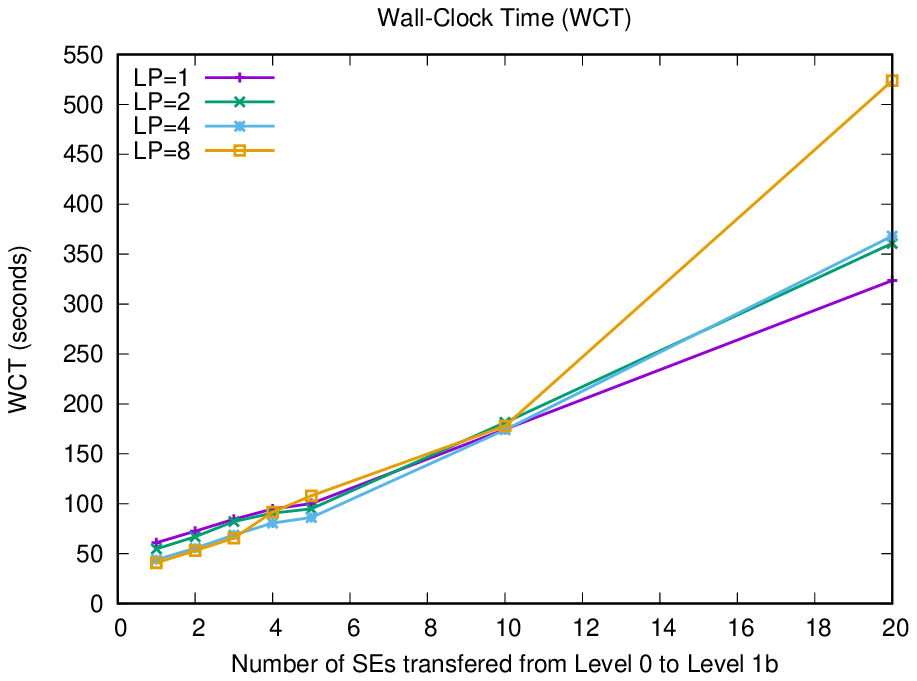}
\caption{Level 0 + Level 1b: Scalability evaluation (WCT) -- increasing number of SEs transferred from Level 0 to Level 1b}
\label{fig_L0L1b}
\end{figure}

In this section, we assess the performance of the combined simulator as composed of the $L_0$ agent-based simulator and $L_{1b}$ (OMNeT++). Both of them are run on the Linux server. In this case, our goal is to study the performance of the simulator with an increasing number of SEs that are transferred from the Level 0 simulator to the Level 1b. Figure~\ref{fig_L0L1b} shows the WCT required to complete a sequential simulation (LP=1) with 16000 SEs when a single $L_{1b}$ spawn is triggered. The lines for LP={2,4,8} refer to simulations in which every LP triggers a single $L_{1b}$ spawn that is run in parallel with the others $L_{1b}$ instances. The WCT increases with the number of SEs handled by the Level 1. Such increment is linear but significant and it is manly due to the sequential architecture of OMNeT++. This confirms our design in which the part of the SEs are handled at Level 0 and costly Level 1 simulations are triggered only when strictly necessary. 

In presence of more than 1 LP at Level 0 (i.e. a parallel Level 0 simulation), each LP is able to trigger its own Level 1 spawns. The cost of such spawns, when they run concurrently, is quite limited, in fact, they can be run in parallel. When both the number of LP is high (e.g. LP=8) and the number of transferred SEs is larger than 8 then the WCT sharply increases. This is mainly due to the amount of memory that is consumed by the Level 0 instances. 
Consequently, the memory use is so high to cause a virtual memory thrashing effect.
When the model requires many Level 1 instances, that are  also quite populated, then distributed setups should be explored, in which the Level 1 instances are partitioned on a set of interconnected hosts.

\subsection{Level 0, Level 1a and Level 1b}
In the last part of this section, we assess the performance of the whole distributed simulator as composed of the $L_0$ agent-based simulator, $L_{1a}$ (ADVISOR - MATLAB/Simulink) and $L_{1b}$ (OMNeT++). It is worth nothing that $L_0$ and $L_{1b}$ are run on the same Linux server, while $L_{1a}$ is run on an interconnected Windows PC. In this performance evaluation, the number of SEs that is transferred from the Level 0 simulator to the Level 1 simulators is set to 1. As demonstrated above, increasing the number of SEs would increase the WCT required to complete the simulation runs. In this part of the performance evaluation, we are more interested in evaluating the cost of triggering and coordinating many instances of different simulators.

%
%
\begin{figure}[h]
\centering
\includegraphics[width=8.5cm]{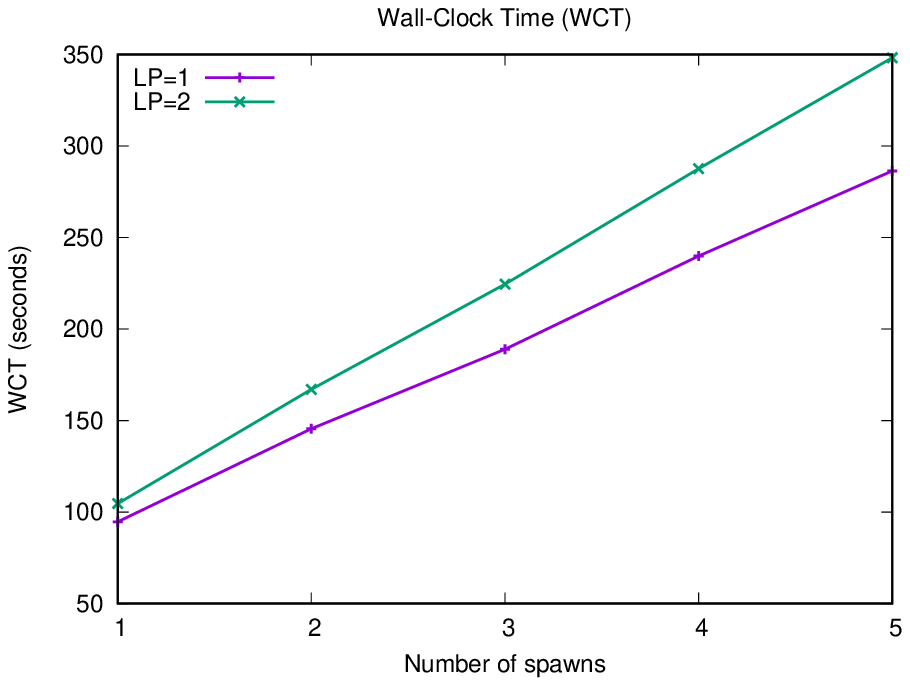}
\caption{Level 0 + Level 1a + Level 1b: Scalability evaluation (WCT) -- 16000 SEs with an increasing number of $L_{1a}$ and $L_{1b}$ spawns.}
\label{fig_L0L1ab-wct}
\end{figure}

Figure~\ref{fig_L0L1ab-wct} shows the amount of time required to complete a sequential simulation (LP=1) in which 16000 SEs are managed at Level 0 while an increasing number of sequential (equally spaced) $L_{1a}$ and $L_{1b}$ spawns are triggered. As reported before, the number of SEs transferred from Level 0 to the Level 1 is set to 1.
The increase of the WCT is linear with the number of instances that are triggered. The cost of triggering a combined Level 1a and 1b instance is quite relevant, since the execution of these simulators is sequential and both of them are monolithic. With LP=2, we mean a parallel $L_0$ simulation in which both the LPs, at given points in time, trigger $L_{1a}$ and $L_{1b}$ spawns. This means that, the amount of work, with respect to LP=1, is much higher but it is (possibly) balanced by the parallelization of the Level 0 model given by the parallel setup. In this case, the pairs of Level 1a+1b spawns that are triggered by the LPs can be executed in parallel and therefore the impact on the WCT is quite limited.

\begin{figure}[h]
\centering
\includegraphics[width=8.5cm]{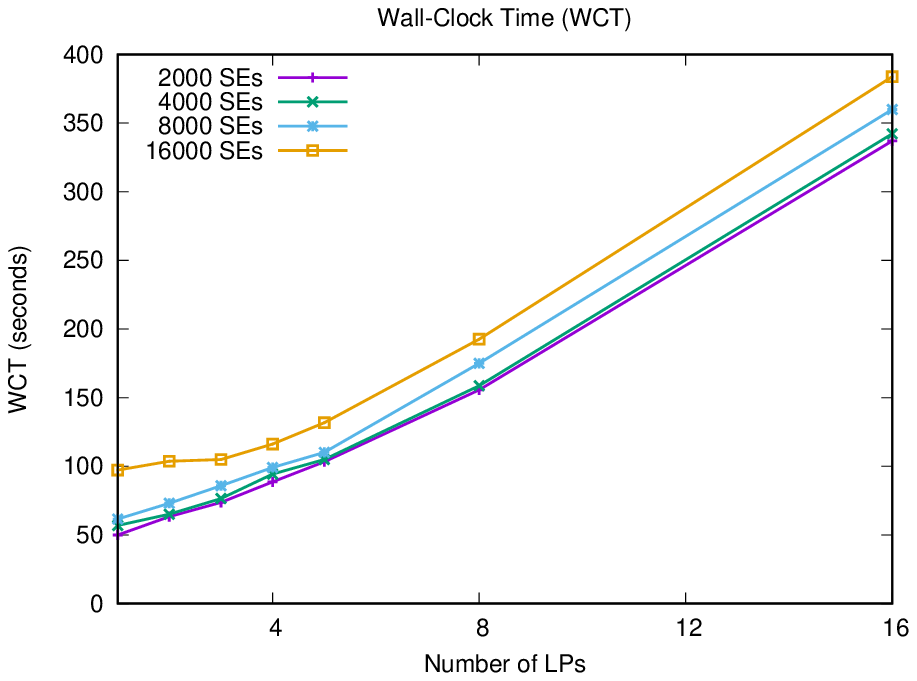}
\caption{Level 0 + Level 1a + Level 1b: Scalability evaluation (WCT) -- 16000 SEs with an increasing number of LPs, each LP triggers a single Level 1a + 1b spawn.}
\label{fig_L0L1ab-wct-LPs}
\end{figure}

Finally, in Figure~\ref{fig_L0L1ab-wct-LPs} the \{2000, 4000, 8000, 16000\} SEs are partitioned in an increasing number of LPs. In this case, at the same simulated time each LP triggers a single Level 1a + 1b spawn. This means that, with LP=16, a total amount of 16 concurrent Level 1a+1b instances are executed. The figure shows that, for LP$\leq$16 the overhead introduced by the many concurrent instances of Level 1 simulators is linear. Further increasing the number of LPs causes memory thrashing on the host running the $L_{1b}$ spawns. As expected, the number of SEs that are partitioned among the LPs has an impact on the WCT but this is limited with respect to increasing the number of LPs. In fact, the efficiency of the $L_0$ simulator is better than the Level 1a + 1b spawns. In our view, this demonstrates that the multi-level execution of hybrid simulators is a viable approach also under the performance viewpoint. More specifically, in the case that a large number of Level 1b spawns needs to be concurrently executed then multiple execution units will be needed to reduce the overhead introduced by this simulation component (e.g.~running in parallel the Level 1b spawns on different hosts).

\subsection{Discussion}
The use of multi-level simulation allows scaling up to higher numbers of SEs, with respect to the use of a single fine-grained simulator, which captures all the complex technical details of the SEs, and their interactions. 
A hybrid simulation approach allows mimicking all these details only when needed. Hence, during the rest of the simulation, the hybrid simulator behaves as a coarse-grained simulator.

As concerns the considered simulation strategies employed in the performance evaluation, in this work we did not assessed the behavior of the adaptive distribution and migration capabilities of the employed agent-based Level 0 simulator~\cite{gda-simpat-2014}. Indeed, evaluating such strategies, when combined with hybrid architectures, might result as a complicated task. Since the goal of this work was to investigate the viability of the use of a hybrid simulation architecture, we did not considered these aspects. Nevertheless, the benefits on the use of adaptive migration strategies in PADS environments have been already shown in the literature~\cite{gda-simpat-2017-part}. The task of properly assessing a whole hybrid simulation with adaptive PADS is an interesting aspect, and it is left as a future work.

Finally, as regards to the simulated application scenario, results suggest that smart dissemination strategies are needed to effectively distribute messages in complex networks, such as those modeling an IoT scenario. A naive gossip strategy, in fact, is not sufficient when a high amount of messages are generated by a large number of devices. Indeed, the amount of received data to be relayed might overwhelm the (in certain cases, poor) communication and computing infrastructures of certain devices.

%% file: sec_conclusions.tex
\section{Conclusions}
\label{sec:conc}

In this paper, we discussed some main issues to cope with, in order to properly simulate the IoT. Scalability and high level of details are the two main, and quite often counterposed, goals. We overviewed some existing techniques, reaching the conclusion that the use of hybrid simulation is a good strategy to employ in this context. 
In substance, this methodology enables the combination of multiple modalities of simulation. This way, complex scenarios can be decomposed into simpler ones, each one being simulated through a specific simulation strategy. 
This approach fosters the re-use of existing simulation components, such as models and simulators, even when these work over different platforms, architectures and operating systems.
All the employed simulation building blocks need to be synchronized and coordinated. If we are able to do it, then this simulation methodology is an ideal one to represent IoT scenarios, which are usually very demanding, due to the heterogeneity of possible scenarios arising from the massive deployment of an enormous amount of sensors and devices.

The analysis of the use case, related to the design of smart services for smart cities and decentralized areas, shows that hybrid simulation techniques provide means to simulate wide geographical areas, with a multitude of simulation entities (agents). However, when needed it is possible to trigger more detailed, fine grained simulations, so as to consider aspects which could not be simulated otherwise. The interesting aspect of this approach is that the detailed (and more costly) simulation can be performed in a specific, limited simulated area, only for the needed time interval of the simulation.

In particular, we have shown an example of a distributed hybrid simulation tool that is composed of three different simulation models, namely, an adaptive agent-based PADS, an OMNeT++ based discrete event simulator and a script-language simulator based on MATLAB. 
These three simulators act in very different ways and run on different hosts. We discussed how it is possible to orchestrate their execution so as to build a sophisticated hybrid simulation of the considered IoT application scenario.
Results from the conducted performance analysis confirmed the viability of the proposed approach. 

As concerns the ongoing activity on the proposed hybrid simulator, built for the simulation of the smart shires use case, we are investigating the use of virtualization technologies. In particular, our goal is to
incapsulate each simulation component in a different virtual machine, to further foster component usability and independence.